\documentclass[prd,preprintnumbers,floatfix,nofootinbib,superscriptaddress, twocolumn]{revtex4}

\usepackage{amsmath,amssymb,hyperref,subfigure,xspace}
\usepackage{mciteplus,color,xcolor,mathtools,graphicx}
\usepackage{braket,bbm,slashed, bm}
\usepackage{multirow}
\usepackage{placeins}
\usepackage[english]{babel}
\usepackage{mathrsfs}
\usepackage{enumerate}
\usepackage{xspace}
\usepackage{slashed}
\usepackage[utf8]{inputenc}
\usepackage[T1]{fontenc}

\usepackage[letterspace=-10]{microtype} 

\newcommand{\Ac}{\mathcal{A}}

\newcommand{\Fc}{\mathcal{F}}

\newcommand{\Hc}{\mathcal{H}}

\newcommand{\Kc}{\mathcal{K}}

\newcommand{\Mc}{\mathcal{M}}

\newcommand{\beq}{\begin{equation}}
\newcommand{\eeq}{\end{equation}}

\newcommand{\kev}{\ensuremath{{\mathrm{\,ke\kern -0.1em V}}}\xspace}
\newcommand{\mev}{\ensuremath{{\mathrm{\,Me\kern -0.1em V}}}\xspace}
\newcommand{\gev}{\ensuremath{{\mathrm{\,Ge\kern -0.1em V}}}\xspace}
\newcommand{\tev}{\ensuremath{{\mathrm{\,Te\kern -0.1em V}}}\xspace}

\usepackage{mfirstuc} % to uppercase the name
\newcommand{\addReviewer}[2]{
  \expandafter\newcommand\csname #1\endcsname[1]{{\bf \color{#2} \capitalisewords{#1}:\,##1}}
  \expandafter\newcommand\csname #1cor\endcsname[2]{{\color{#2} \capitalisewords{#1}:\,\st{##1}{\bf ##2}}}
  \expandafter\newcommand\csname #1color\endcsname{#2}
}

\usepackage{soul,color}
\definecolor{chromeyellow}{rgb}{1.0, 0.65, 0.0}
\definecolor{DodgeBlue}{rgb}{0.118, 0.565,1.000}
\definecolor{asparagus}{rgb}{0.53, 0.66, 0.42}
\definecolor{cadmiumgreen}{rgb}{0.0, 0.42, 0.24}

\addReviewer{andrew}{cadmiumgreen}

\hypersetup{ 
    pdfnewwindow=true,      % links in new window
    colorlinks=true,       % false: boxed links; true: colored links
    linkcolor=blue,          % color of internal links
    citecolor=blue,        % color of links to bibliography
    filecolor=blue,      % color of file links
    urlcolor=blue        % color of external links
} 

\definecolor{jlab_red}{RGB}{192,39,45}
\definecolor{jlab_orange}{RGB}{249,102,0}
\definecolor{jlab_blue}{RGB}{47,122,121}
\definecolor{jlab_green}{RGB}{65,125,10}

\begin{document}

%%%%%%%%%%%%%%%%%%%%%%%%%%%%%%%%%%%%
% Title
%%%%%%%%%%%%%%%%%%%%%%%%%%%%%%%%%%%%
\title{ Constraining $1+\mathcal{J}\to 2$ coupled-channel amplitudes in finite-volume }

%%%%%%%%%%%%%%%%%%%%%%%%%%%%%%%%%%%%
% Author list
%%%%%%%%%%%%%%%%%%%%%%%%%%%%%%%%%%%%
%%%%%%%%%%
\author{Ra\'ul A. Brice\~no}
\email[e-mail: ]{rbriceno@jlab.org}
\affiliation{\lsstyle Thomas Jefferson National Accelerator Facility, 12000 Jefferson Avenue, Newport News, VA 23606, USA}
\affiliation{ Department of Physics, Old Dominion University, Norfolk, Virginia 23529, USA}
%%%%%%%%%%

%%%%%%%%%%
\author{Jozef J. Dudek}
\email[e-mail: ]{dudek@jlab.org}
\affiliation{\lsstyle Thomas Jefferson National Accelerator Facility, 12000 Jefferson Avenue, Newport News, VA 23606, USA}
\affiliation{ Department of Physics, College of William and Mary, Williamsburg, VA 23187, USA}

%%%%%%%%%%
\author{Luka Leskovec}
\email[e-mail: ]{leskovec@jlab.org}
\affiliation{\lsstyle Thomas Jefferson National Accelerator Facility, 12000 Jefferson Avenue, Newport News, VA 23606, USA}
\affiliation{ Department of Physics, Old Dominion University, Norfolk, Virginia 23529, USA}
%%%%%%%%%%

%%%%%%%%%%%%%%%%%%%%%%%%%%%%%%%%%%%%
% Preprint Number
%%%%%%%%%%%%%%%%%%%%%%%%%%%%%%%%%%%%
\preprint{JLAB-THY-21-3365}

%%%%%%%%%%%%%%%%%%%%%%%%%%%%%%%%%%%%
% Abstract  & PACS
%%%%%%%%%%%%%%%%%%%%%%%%%%%%%%%%%%%%
\begin{abstract}
Whether one is interested in accessing the excited spectrum of hadrons or testing the standard model of particle physics, electroweak transition processes involving multi-hadron channels in the final state play an important role in a variety of experiments. Presently the primary theoretical tool with  which one can study such reactions is lattice QCD, which is defined in a finite spacetime volume. In this work, we investigate the feasibility of implementing existing finite-volume formalism in realistic lattice QCD calculation of reactions in which a stable hadron can transition to one of several two-hadron channels under the action of an external current. We provide a conceptual description of the \emph{coupled-channel transition} formalism, a practical roadmap for carrying out a calculation, and an illustration of the approach using synthetic data for two non-trivial resonant toy models. The results provide a proof-of-principle that such reactions can indeed be constrained using modern-day lattice QCD calculations, motivating explicit computation in the near future. 
\end{abstract}
\date{\today}
\maketitle

%%%%%%%%%%%%%%%%%%%%%%%%%%%%%%%%%%%%%%%%%%%%%%%%%%%%%%%%%%%%%%%%%%%%%%%%%%%%%%%%%%%%%%%%%%%%%%%%%%%%%%%%%%%%%%%%%%%%%%%
%%%%%%%%%%%%%%%%%%%%%%%%%%%%%%%%%%%%%%%%%%%%%%%%%%%%%%%%%%%%%%%%%%%%%%%%%%%%%%%%%%%%%%%%%%%%%%%%%%%%%%%%%%%%%%%%%%%%%%%
%%%%%%%%%%%%%%%%%%%%%%%%%%%%%%%%%%%%%%%%%%%%%%%%%%%%%%%%%%%%%%%%%%%%%%%%%%%%%%%%%%%%%%%%%%%%%%%%%%%%%%%%%%%%%%%%%%%%%%%

%%%%%%%%%%%%%%%%%%%%%%%%%%%%%%%%%%%%
% Section I :: introduction
%%%%%%%%%%%%%%%%%%%%%%%%%%%%%%%%%%%%
\section{Introduction}\label{sec:I}

A primary mechanism for producing the excited resonances of Quantum Chromodynamics (QCD) is a high-energy reaction where a current lying outside of the strong interaction sector interacts with a stable hadron. 
When the produced hadron resonance then decays into a pair of stable hadrons, we refer to this as a $1 \xrightarrow{\mathcal{J}} 2$ process, and we would like to be able to describe such processes from first principles within QCD. A particularly important case, which we examine in this paper, has the resonance able to decay into more than one final-state hadron-hadron system, that is to say a \emph{coupled-channel} situation. 
A contemporary experimental example might be the claim of pentaquark resonances in the $J/\psi \, p$ final state, produced in $\Lambda_b$ hadronic decays~\cite{Aaij:2015tga,Aaij:2019vzc}, which can in principle also be produced in photoproduction off a proton, $\gamma p \to J/\psi p$~\cite{Ali:2019lzf}. The resonances observed lie in an energy region where coupled channels like $\overline{D} \Sigma_c$ open up, necessitating a coupled-channel approach in photoproduction. Other potential applications include semileptonic weak decays of $b$-- or $c$--quark containing hadrons which typically have sufficient energy to access more than one hadron-hadron final-state~\cite{Meissner:2013pba, Gambino:2020jvv}.

Amplitudes describing $1 \xrightarrow{\mathcal{J}} 2$ processes potentially allow access to information that may shed light on the internal structure of resonances. With an explicit parameterization of the transition amplitude it is possible to analytically continue to the complex values of the hadron-hadron energy. If the system resonates we expect a pole singularity, and the residue of that pole has an interpretation in terms of the transition form-factor of the resonance. By examining the dependence of this form-factor on the current virtuality, it may be possible to infer information about the spatial distribution of constituents of the resonance.

%%%%%%%%%%%%%%%%%%%%%%%%%%%%%%%%%%%%%%%%%%%%%%%%%%%%%%%%%%%%%%

Lattice QCD offers the only first-principles approach to computation of transition matrix elements within QCD, but by necessity, these calculations are done in a \emph{finite spatial volume}, and this leads to effects which are in general not simply small corrections to infinite-volume quantities. The finite-volume effects are in fact dominant features that need to be handled carefully using a rigorous formalism. It is these effects that we will explore in this paper.

In order to study processes in which one or several hadron-hadron final states are produced when a current is absorbed by a stable hadron, a two-stage lattice QCD computation is required. In the first stage, the hadron-hadron scattering matrix (without involvement of the current) is determined. This is an increasingly common calculation in which the finite-volume spectrum is extracted from a matrix of two-point correlation functions. A `diagonalization' of this matrix through solution of a generalized eigenvalue problem leads to the spectrum (from the eigenvalues) and `optimal' operators for each energy level as a linear combination of the basis operators (through the eigenvectors)~\cite{Michael:1985ne, Luscher:1990ck, Blossier:2009kd, Dudek:2007wv}.
The energy levels provide constraint on the scattering matrix~\cite{He:2005ey, Hansen:2012tf, Briceno:2012yi, Briceno:2014oea}, but in practice exploiting the relationship between the finite-volume spectrum and the desired scattering amplitudes requires resorting to construction of energy-dependent parametrizations for the amplitudes~\cite{Guo:2012hv}. More constraint is provided by having more energy levels, and this can be achieved by computing in multiple lattice volumes and/or by considering hadron-hadron systems with net momentum (`in-flight'). This methodology has proven useful and has allowed for the determination of numerous coupled-channel scattering amplitudes via lattice QCD~\cite{Dudek:2014qha, Wilson:2014cna, Briceno:2017qmb, Dudek:2016cru, Woss:2019hse, Moir:2016srx, Woss:2020ayi}. For a recent introductory review on these concepts see Ref.~\cite{Briceno:2017max}.

The second stage, in which the current is introduced, involves computation of three-point correlation functions with source and sink hadronic operators having definite values of three-momentum allowed on a cubic lattice. The source operator is chosen to be one which interpolates the initial state stable hadron, and the sink operator is ideally constructed to interpolate only a single one of the finite-volume eigenstates having the quantum numbers of the final state hadron-hadron system. A powerful approach to achieve this is to make use of the `optimal' operators found in the two-point variational calculation described above~\cite{Shultz:2015pfa, Dudek:2009kk, Becirevic:2014rda}. At intermediate times between the source and sink, the relevant current is inserted, and by analyzing the insertion time dependence, a matrix element describing the transition at one particular kinematic point can be extracted.
 
Through the use of multiple three-momenta at source and sink, and by considering finite-volume energy levels which span a range of energies, we can obtain a sampling of both the current-virtuality and scattering energy dependence of the transition process, but only once the effects of the finite-volume are accounted for. 

The relevant formalism to map finite-volume matrix-elements into infinite-volume matrix-elements already exists. References~\cite{Briceno:2014uqa, Briceno:2015csa} generalize the original idea of the ``Lellouch-L\"uscher'' factor~\cite{Lellouch:2000pv} to be applicable in essentially all cases of $1 \xrightarrow{\mathcal{J}} 2$ transition amplitudes. The formalism has been applied to explicit lattice QCD results so far only in the case where elastic $\pi\pi$  is the sole relevant scattering channel. 
First implementations of the Lellouch-L\"uscher formalism were performed in $K\to \pi\pi$ weak decay studies~\cite{Bai:2015nea,Blum:2012uk,Blum:2011pu,Blum:2011ng}, where the physical process is constrained to lie at only a single kinematic point. 
References~\cite{Briceno:2016kkp, Briceno:2015dca, Alexandrou:2018jbt} considered $\gamma \pi \to \pi\pi$ with $J^P=1^-$, extracting the infinite-volume transition matrix elements as a function of $E_{\pi\pi}$ and photon virtuality. By analytically continuing to the $\rho$ resonance pole at a complex value of $E_{\pi\pi}$, the \emph{resonance transition form-factor} for $\rho \to \pi \gamma$ was determined. In this paper we will explore how calculations like these can be extended into the coupled-channel sector, where a resonance might have decays to more than one hadron-hadron final state.

The formalism for the coupled-channel case is already laid out~\cite{Briceno:2014uqa, Briceno:2015csa}, but in this paper we will present some observations which provide a conceptual interpretation of the approach, as well as providing a proposal for practical implementation. Our aim is to investigate whether contemporary lattice QCD calculations can obtain sufficient constraint so that coupled-channel transition amplitudes can be reliably determined. Our exploration will come in the form of toy-models of two-channel resonant scattering. We will propose explicit scattering and transition amplitudes, and from them generate synthetic finite-volume spectra and matrix-element data simulating the situation in a typical lattice QCD calculation. We will then analyze the synthetic data using the finite-volume techniques to establish that the essential features of the original amplitudes can be reproduced. This procedure parallels that presented in Ref.~\cite{Guo:2012hv} for analysis of coupled-channel finite-volume spectra, which proved to be a realistic prediction of how explicit lattice QCD calculations would later be handled~\cite{Dudek:2014qha, Wilson:2014cna, Briceno:2017qmb, Dudek:2016cru, Woss:2019hse, Moir:2016srx, Woss:2020ayi}. 

In this paper, we consider the simplest non-trivial scenario for $1 \xrightarrow{\mathcal{J}} 2$ reactions. We work in a situation where the scattering channels are composed of identical spinless particles, and for the kinematics considered, up to two channels will be open which are completely saturated by the $\ell=0$ partial wave. We will consider toy models in which a single resonance couples to both decay channels -- we focus our attention on resonant systems not just because of their physical interest, but also because they are potentially challenging owing to their rapid energy-dependence. If the procedure to be presented is successful for resonant processes, non-resonant systems should not present difficulties.

In Section II we discuss how $1 \xrightarrow{\mathcal{J}} 2$ coupled-channel processes are described in infinite and finite volumes, before presenting in Section III a pair of relevant toy-models and their finite-volume spectra and matrix-elements. In Section IV we generate synthetic data for these toy models which resemble those accessible in contemporary lattice QCD calculations, and show how starting with these data we would go about determining the infinite-volume transition matrix elements. In Section V we summarize, and make projections for likely explicit lattice QCD applications of the methodology we have laid out. 
In appendices we provide some supporting illustrations of properties of the Lellouch-L\"uscher factor in several cases including those where a channel is kinematically closed, where multiple partial-waves must be considered, and where a bound-state appears far below any decay channel. 
A discussion of the possibility of defining a version of the Lellouch-L\"uscher factor at energies away from solutions of the finite-volume quantization condition is presented, where we conclude that this cannot be done in a unique way.
We also illustrate the rather unique properties of one of the simplest parameterizations of a coupled-channel resonance, the \emph{Flatt\'e} amplitude, presenting some caveats regarding the flexibility of such a form in the description of scattering and transition amplitudes.

%%%%%%%%%%%%%%%%%%%%%%%%%%%%%%%%%%%%
% Section II :: formalism
%%%%%%%%%%%%%%%%%%%%%%%%%%%%%%%%%%%%
\vspace{-6mm}
\section{$1\to2$ processes in infinite and finite volume}\label{sec:II}
\vspace{-3mm}

We will concern ourselves with processes in which a current acting upon an initial single-hadron state induces a transition into a hadron-hadron sector, where strong rescattering may occur. In the cases we wish to focus upon, the $S$-matrix describing the rescattering in a given partial-wave features multiple kinematically accessible hadron-hadron channels, a situation described as \emph{coupled-channel} scattering. We will consider only cases in which the scattering particles carry no intrinsic spin, although removing this restriction is quite straightforward~\cite{Briceno:2014oea, Briceno:2013hya}. We begin by specifying the relevant kinematics and properties of the hadron-hadron scattering amplitudes.

%%%%%%%%%%%%%%%%%%%%%%%%%%%%%%%%%%%%%%%%%%%%%%%%%%%%%%%%%
\subsection{$2\to2$ scattering in infinite volume} \label{sec:IVamps}
%%%%%%%%%%%%%%%%%%%%%%%%%%%%%%%%%%%%%%%%%%%%%%%%%%%%%%%%%

We will mostly follow the notation used in Refs.~\cite{Briceno:2014uqa, Briceno:2015csa}, where indices $a,b$ label hadron-hadron channels, and $\ell$ and $m_\ell$ refer to the total and azimuthal components of the angular momentum for a system projected into a definite partial-wave. The two hadrons in each channel do not need to be identical, but for simplicity in this study they will be, with mass $m_{a}$ in channel $a$.
Quantities measured in the center-of-momentum frame of the hadron-hadron system are labeled by a star, e.g.
\begin{equation}
  q_a^\star = \tfrac{1}{2} \sqrt{ s - 4\,m_{a}^2   }, \label{qcm}
\end{equation}
is the relative momentum, where $s= E^{\star \,2}$. The corresponding \emph{phase-space} for identical particles\footnote{an additional factor of 2 appears in the case of non-identical particles} is
\begin{equation}
\rho_a(s) =    \, \frac{ q^{\star}_a}{16\pi \sqrt{s}}\, ,
\label{eq:ps}
\end{equation}
and above $N$ open thresholds, it is convenient to introduce a matrix $\rho$, defined as
\begin{equation*}
\rho ={\rm diag}(\rho_1,\rho_2,\ldots,\rho_N).
\end{equation*}

The $S$--matrix and the scattering amplitude, $\Mc(s)$, are related in general via,
\begin{equation*}
  S = 1 + 2i\,\sqrt{\rho}\,\Mc \,\sqrt{\rho } \,,
\end{equation*}
and taking advantage of rotational symmetry in an infinite volume, the scattering amplitude can be partial-wave expanded and the resulting independent partial-wave amplitudes labeled by $\ell$. In particular, we label the scattering amplitude coupling channels $a$ and $b$ in the $\ell$ partial wave as $\Mc_{ab,\ell}(s)$. Time-reversal symmetry ensures that $\Mc$ is a symmetric matrix in channel space.

\emph{Unitarity} in the coupled-channel case can be expressed as a constraint on the imaginary part of the matrix inverse of $\Mc_\ell$,
\begin{equation}
\mathrm{Im}\, \big[\Mc_\ell^{-1}(s) \big]_{ab} = - \delta_{ab} \, \rho_a(s) \, \Theta( s - s^\mathrm{thr}_a),
\label{eq:unitarity}
\end{equation}
where the step-function ensures that the imaginary part is zero below the kinematic threshold, ${\sqrt{s^\mathrm{thr}_a} = 2\,m_{a} }$.

While elements of the scattering matrix can be obtained physically only for real values of $s$ above kinematic thresholds, it proves relevant to consider amplitudes more generally as functions of complex $s$, and in particular to pay attention to their singularities.  Unitarity and the presence of the square-root in Eq.~\eqref{qcm} ensure the scattering matrix has branch point singularities at each kinematic threshold, and this renders $\Mc(s)$ a multivalued complex function which can be described by a Riemann sheet structure. In addition to branch cuts coming from unitarity, \emph{pole singularities} can be present and depending upon their location, they can be interpreted as representing stable bound-states or unstable resonances. Near to a resonance pole located at $\sqrt{s_R}=m_R\pm i\tfrac{1}{2}\Gamma$, the scattering matrix takes the form,
\begin{align}
\label{eq:poledef}
\Mc_{ab}(s)
\sim \frac{c_a\,c_b}{s_R-s},
\end{align}
where the complex-valued $c_a$ is interpreted as the coupling of the resonance to channel $a$. In this expression we have left the angular momentum dependence implicit, but for spinless scattering particles, a resonance only appears for a single value of $\ell$, fixed by the angular momentum of the resonance. These resonance poles are objects of central interest in the study of hadron spectroscopy\footnote{Other singularities can also be present in $\Mc(s)$ but are usually of lesser importance in determining resonance properties: projection into partial-waves obscures the role of crossing symmetry, such that unitarity in the crossed-channels leads to cut-like singularities, known as \emph{left-hand cuts} which appear in the partial-wave projected amplitudes for values of $s$ typically well below kinematic thresholds.}.

%%%%%%%%%%%%%%%%%%%%%%%%%%%%%%%%%%%%%%%%%%%%%%%%%%%%%%%%%%%%%%%%%%%%%%%%%%%%%%%%%%%%%%%
\subsection{$1\to 2$ transition amplitudes in infinite-volume \label{sec:transamps}}

We consider the process in which a single stable hadron of mass $M$ with four-momentum $P_i$ is acted upon by a current to become a hadron-hadron system with four-momentum $P_{\!f}$ and invariant-mass squared $s = P_{\!f}^{\,2}$. We will adopt a notation where $1 \xrightarrow{\mathcal{J}} 2$ amplitudes, $\Hc^\mu$, carry an explicit Lorentz index, motivated by the most likely application to vector or axial-vector currents, however the bulk of our results will be applicable to a more general class of currents. Since the amplitude describes a situation where a particular hadron-hadron channel is produced in the final-state, the amplitude will carry a single channel label, $\Hc^\mu_a$. Unlike the amplitudes describing $2 \to 2$ scattering, because the current has non-trivial rotational properties, the partial-wave projected transition amplitudes in general depend on the azimuthal component, $m_\ell$. A somewhat over-complete notation featuring the current virtuality, $Q^2\equiv  -(P_f-P_i)^2$,  $\Hc^\mu_{a, \ell m_\ell}( P_f, P_i;\, Q^2, s)$, will often be simplified by suppressing arguments or indices irrelevant to the discussion at that point. 

In terms of matrix elements of local currents in an infinite-volume, these amplitudes can be expressed as, 
\begin{equation}
\label{eq:inf_ME}
\Hc^\mu_{a, \ell m_\ell}( P_f, P_i;\, Q^2, s) = 
 {}_\infty\langle {P}_{\!f}; a \ell m_\ell \big| \mathcal{J}^{\mu}(x\!=\!0) \big| \mathbf{P}_{\!i} \rangle_\infty, 
\end{equation}
where the argument of the current, $x\!=\!0$, is introduced to emphasized that the current is evaluated at the origin.
The initial state $\big| \mathbf{P}_{\!i} \rangle_\infty$, is an infinite-volume single-particle on-shell state which has the standard relativistic normalization. Because the state is on-shell, its energy is completely determined from its spatial momentum, $\mathbf{P}_{\!i}$, and the mass of the particle. 
The final state, $ {}_\infty \langle {P}_{\!f}; a \ell m_\ell \big|$, is a two-particle state in channel $a$ with definite angular momentum $(\ell, m_\ell)$. In the center-of-momentum frame, one can construct such a state by partial-wave projecting products of single-particle states, which can then be boosted to an arbitrary frame (see, for example, Appendix D in Ref.~\cite{Briceno:2016kkp}). Because the energy of a two-particle state is not fixed by its spatial momenta, these states in general must be labeled by their four-momenta, ${P}_{\!f}$, where ${ {P}_{\!f}^2 = (E^\star)^2 = s }$.
~\footnote{From Eq.~\eqref{eq:inf_ME}, one can verify the dimensions of the transition amplitude, $\Hc^\mu$. The relativistic single-particle states have energy dimensions of $[E]^{-1}$, and consequently the two-particle states have dimensions $[E]^{-2}$. If we assume the current to be the electromagnetic current, the dimensions of the current are $[E]^{-3}$, and consequently the amplitude is dimensionless.
}

Unitarity provides a constraint on transition amplitudes which can be satisfied by expressing them as~\cite{Briceno:2014uqa, Briceno:2020vgp},
\begin{align}
\label{eq:trans_amp}
\Hc^{\mu}_{a,\ell m_\ell} &(P_f, P_i;Q^2,s) = \cr
&\sum_b
\Ac^{\mu}_{b,\ell m_\ell}(P_f, P_i;Q^2,s)
\,
\frac{1}{q_b^{\star \ell} }
\,
\Mc_{ba,\ell}(s),
\end{align}
where the unitarity branch cuts required to be present in $\Hc$ are housed in $\Mc$. The function $\Ac$ which depends both on the virtuality $Q^2$ and the invariant mass $s$, does not have unitarity branch cuts in $s$ and as such $\Ac$ should be a real smooth function of $s$ above thresholds~\footnote{Note that Eq.~\eqref{eq:trans_amp} is not a unique solution to the unitarity condition. An alternative solution, commonly applied in the case of elastic scattering is the Omn\`es solution~\cite{Omnes:1958hv}, constructed from a dispersive integral featuring the scattering phase. This solution has the same imaginary part, but differs in the real part, and the net difference between our choice and this alternative is the particular form of the smooth function multiplying it. }. 
The factor $\frac{1}{q_b^{\star \ell} }$ is required to deal with the mismatch between the threshold behaviors of $\Hc$ and $\Mc$, where $\Mc_{ab, \ell} \sim q_a^{\star\ell} \, q_b^{\star\ell}$ while $\Hc_{a, \ell m_\ell} \sim q_a^{\star\ell}$.

While $\Ac$ lacks $s$-channel singularities in $s$, it can have singularities in the virtuality variable $Q^2$. One example is when $Q^2$ has a timelike value large enough to produce a pair of hadrons there will be a branch point reflecting unitarity in the scattering of those two particles. In this paper we will not concern ourselves with analytic structure in $Q^2$ as in our eventual lattice QCD application we will typically be restricted to values of $Q^2$ away from such singularities. 

\vspace{5mm}

The $\ell, m_\ell$ subscript on $\Hc$ indicates that our approach is to first perform a partial-wave projection of the final hadron-hadron system, and to then express the dependence on the Lorentz structure of the current in terms of Lorentz-covariant kinematic structures and Lorentz-invariant amplitudes for a final state of angular-momentum $\ell$. A simple example illustrates the method, and provides a physically relevant case for later study: We will consider stable scalar hadrons labeled $\chi, \varphi_1, \varphi_2$, and a conserved vector current (whose quanta are labeled $\gamma$), working in an energy region where the processes $\chi \gamma \to \varphi_1 \varphi_1$ and $\chi \gamma \to \varphi_2 \varphi_2$ are kinematically allowed. We will focus on $S$-wave ($\ell=0$) scattering in the hadron-hadron channels such that Eq.~\eqref{eq:trans_amp} becomes
\begin{equation*}
 \Hc^\mu_{a, 00} = \sum\nolimits_b \Ac^\mu_{b,00} \, \Mc_{ba,0}\, ,
\end{equation*}
where $\Mc$ is the $S$-wave scattering matrix in the $(\varphi_1 \varphi_1, \varphi_2 \varphi_2)$ channel space. Considering the first element, $\Ac^\mu_{1, 00}$, describing primary production of the $\varphi_1 \varphi_1$ channel, we can perform a Lorentz decomposition using the fact that the initial and final states are both scalars to write,
\begin{align*}
 \Ac&^\mu_{1, 00}(P_f, P_i; Q^2, s) =\\
 & (P_i+ P_f)^\mu \, \Fc_1(Q^2, s) + (P_f - P_i)^\mu \, \mathcal{G}_1(Q^2, s) \, ,
\end{align*}
which simplifies when conservation of the vector current is applied, to give,
\begin{align*}
 \Ac&^\mu_{1, 00}(P_f, P_i; Q^2, s)=\cr
 & \left[ (P_i+ P_f)^\mu  + \tfrac{s-M^2}{Q^2} (P_f - P_i)^\mu \right] \mathcal{F}_1(Q^2, s) \,.
\end{align*}
In this case there is just a single Lorentz-invariant form-factor $\mathcal{F}_1(Q^2, s)$ multiplied by a kinematic factor. A decomposition of the same form applies to the second channel such that,
\begin{align}
 \label{eq:Amua}
 \Ac&^\mu_{a, 00}(P_f, P_i; Q^2, s)=\cr
 & \left[ (P_i+ P_f)^\mu  + \tfrac{s-M^2}{Q^2} (P_f - P_i)^\mu \right] \mathcal{F}_a(Q^2, s) \,,
\end{align}
which implies~\footnote{For the electromagnetic current, this decomposition would imply that the $\mathcal{F}_b$ have dimensions of $[E]^{-1}$.},
\begin{align} \label{eq:HPFM}
 \Hc&^\mu_{a, 00}(P_f, P_i; Q^2, s) =  \nonumber \\
 & \left[ (P_i+ P_f)^\mu  + \tfrac{s-M^2}{Q^2} (P_f - P_i)^\mu \right] \sum\nolimits_b \mathcal{F}_b(Q^2, s)\,  \Mc_{b a,0} \,.
\end{align}

In a more general case, for each partial-wave $\ell$ we can write a general decomposition as a \emph{sum} of kinematic factors multiplying linearly-independent form-factors. Care is required to account for factors of $q^\star$ that might appear in the Lorentz-covariant kinematic factors so that the correct threshold behavior appears in Eq.~\eqref{eq:trans_amp}. When $\ell >0$, the dependence on $m_\ell$ enters via the presence of final-state polarization tensors, $\epsilon^{\nu \ldots}(P_f, m_\ell)$, in the kinematic factors.

%%%%%%%%%%%%%%%%%%%%%%%%%%%%%%%%%%%%%%%%
\pagebreak %%%%%%%%%%%%%%%%%%%%%%%%%%%%%
%%%%%%%%%%%%%%%%%%%%%%%%%%%%%%%%%%%%%%%%

As was the case for the $2 \to 2$ scattering matrix $\Mc$, the $1 \xrightarrow{\mathcal{J}} 2$ transition amplitude $\Hc$ can be considered to be a function of \emph{complex} $s$, and in particular as indicated by Eq.~\eqref{eq:trans_amp} and the text immediately after, we expect $\Hc$ to have the same pole singularities as $\Mc$, but with different residues, that we can express generically for $s \sim s_R$ as
\begin{equation}
 \Hc_a(s) \sim \frac{ c_a \, f_R(Q^2) }{s_R - s}\, ,
 \label{eq:H_res}
\end{equation}
where $f_R(Q^2)$ has an interpretation as the \emph{transition form-factor of the resonance}. Expressed in terms of the function $\Ac$ we have
\begin{equation} \label{eq:resff}
 f_R(Q^2) = \sum\nolimits_b c_b\, \Ac_b(Q^2, s_R), \, 
\end{equation}
where the possible complex values of $\{ c_b \}$ and $\Ac$ off the real axis make it clear that $f_R(Q^2)$ need not be real valued. Because $\Ac$ lacks the unitarity cut, it is a continuous function in the complex-$s$ plane, and as a result, its analytic continuation to $s_R$ is trivial.

%%%%%%%%%%%%%%%%%%%%%%%%%%%%%%%%%%%%%%%%%%%%%%%%%%%%%%%%%
\subsection{Finite-volume formalism \label{sec:FVformalism}}

The relationship providing the connection between the $2 \to 2$ scattering matrix and the discrete spectrum of energy eigenstates in a finite-volume, $E^\star_n(\mathbf{P},L)$, can be expressed in the form of a single equation~\cite{He:2005ey, Hansen:2012tf, Briceno:2012yi, Briceno:2014oea},
\begin{align}
\label{eq:luscher}
\det\left[  F^{-1}(E^\star,\mathbf{P};L) + \Mc(E^\star) \right] = 0 \, ,
\end{align}
where in general $F$ and $\Mc$ are matrices in the space of scattering channels and partial-waves. The discrete energies, $E^\star_n(\mathbf{P},L)$, correspond to the solutions of this equation. The corresponding energies in the rest-frame of the lattice are trivially obtained from the energies in the center-of-momentum frame, $E_n = \sqrt{ (E_n^\star)^2 + |\mathbf{P}|^2 }$.

Eq.~\eqref{eq:luscher} is a generalization of L\"uscher's original relation between the finite-volume spectrum and the scattering amplitude for energies where a single channel is kinematically open~\cite{Luscher:1990ux}. Projection into irreducible representations of the cubic symmetry relevant to most lattice QCD calculations can be straightforwardly achieved. An increasingly common approach to application of this relation when a set of discrete energy levels have been determined in an explicit lattice QCD calculation, is to propose energy-dependent parameterizations of $\Mc$~\cite{Guo:2012hv}. The free parameters in these forms are then varied, solving Eq.~\eqref{eq:luscher} for a discrete spectrum at each iteration, with a comparison to the computed spectrum performed in the form of a $\chi^2$. Minimization of this $\chi^2$ leads to a best available description of the scattering matrix. This approach is described in some detail in Ref.~\cite{Briceno:2017max}.

In this paper we seek to extend the application of such a finite-volume technique to the case of $1 \xrightarrow{\mathcal{J}} 2$ coupled-channel transition amplitudes, where the new input is a set of matrix-element values extracted from three-point correlation functions computed using lattice QCD. The formalism presented in Refs.~\cite{Briceno:2014uqa, Briceno:2015csa} provides the relationship between current matrix-elements computed in finite-volume and the infinite volume $1 \xrightarrow{\mathcal{J}} 2$ transition amplitudes, $\Hc$, that we introduced in the previous section. 

\vspace{2mm}
The relation takes the form
\begin{equation}
\label{eq:LL_formula}
\Big| \big\langle E_n, \mathbf{P}_{\!f} \big| \mathcal{J}^{\mu}(x\!=\!0) \big| \textbf{P}_{\!i} \big\rangle_L \Big| 
= \tfrac{1}{L^3 \sqrt{2 E_i} \sqrt{2 E_n} }
\Big[ \Hc^{\mu} \!\cdot\! \widetilde{\mathcal{R}}_n  \!\cdot\! \Hc^{\mu} \Big]^{1/2} ,\, 
\end{equation}
where the final state is one of the discrete energy levels of this finite-volume system, having an energy which solves Eq.~\eqref{eq:luscher}, and the initial state is a single on-shell hadron\footnote{The single hadron state in a finite volume will have a mass which is equal to the infinite volume mass up to exponentially small corrections.}. Both finite-volume eigenstates are normalized to unity. The matrix $\widetilde{\mathcal{R}}_n$ sandwiched between $\Hc$ and its transpose is the ``Lellouch-L\"uscher'' factor, introduced in Ref.~\cite{Briceno:2014uqa}, which is the residue of the finite-volume hadron-hadron propagator at the finite-volume energy, $E_n$. $\widetilde{\mathcal{R}}_n$ is a matrix in partial-waves and channels, defined by
\begin{align}
\widetilde{\mathcal{R}}_n(&\mathbf{P}, L) 
\equiv \nonumber \\ 
&2E_n \cdot\!  \lim_{E \to E_n} (E - E_n) \Big( F^{-1}(E^\star, \mathbf{P}; L) + \Mc(E^\star) \Big)^{-1} \, ,
\label{Rn}
\end{align}
where the energies in the numerator are evaluated in the rest frame of the lattice. This matrix, $\widetilde{\mathcal{R}}_n = 2E_n \!\cdot\! \mathcal{R}_n$, where ${\mathcal{R}}_n$ is the more commonly presented object given in, for example, Eq.~(5) of Ref.~\cite{Briceno:2015csa}. 
The prefactor of $2E_n$ is introduced in order for the denominator of Eq.~\eqref{eq:LL_formula}, $L^3 \sqrt{2 E_i} \sqrt{2 E_n}$, to provide a convenient normalization relating single-particle finite-volume states and their infinite-volume counterparts. For single hadron states we have the following relation, 
\begin{equation*}
| \textbf{P}_{\!i} \big\rangle_\infty \sim \sqrt{2E_i L^3}\, | \textbf{P}_{\!i} \big\rangle_L\, ,
\end{equation*} 
where the equivalence indicates that their matrix elements for local currents are the same up to exponentially suppressed corrections. Similarly, were the two-hadron state to couple to a deeply bound state, it would be the case that $\sqrt{2E_n L^3}$ would provide the necessary normalization to relate finite- and infinite-volume matrix elements. 

In the case of elastic scattering with only a single relevant partial-wave, there is a helpful conceptual interpretation of $\sqrt{\widetilde{\mathcal{R}}_n }$ as the \emph{normalization in a finite-volume} of the hadron-hadron state, $\big| E_n\big\rangle_L$, i.e.~\footnote{This expression is conceptually useful in the case of $1 \xrightarrow{\mathcal{J}} 2$ processes, but it is not an identity. For example, it fails to capture $\mathcal{O}(L^{-3})$ corrections present for finite-volume matrix elements associated with $2 \xrightarrow{\mathcal{J}} 2$ reactions~\cite{Baroni:2018iau, Briceno:2015tza, Briceno:2020xxs}. }
\begin{equation}
\sqrt{2 E_nL^3}\; \big|E_n\big\rangle_L \sim \sqrt{ \widetilde{\mathcal{R}}_n} \, \big|\varphi\varphi(E^\star \!=\!E^\star_n)\big\rangle_\infty \, ,
\end{equation}
and it is interesting to consider if such a picture still holds in the coupled-channel case (still assuming dominance of a single partial-wave). An important observation is that the matrix $\widetilde{\mathcal{R}}_n$, whose \emph{dimension} is simply the number of open channels, only has \emph{rank} $=1$ at the energies, $E^\star_n$, which solve the quantization condition, Eq.~\eqref{eq:luscher} ~\cite{Briceno:2014uqa}. The reduction in rank can be seen by exploring the eigenvector decomposition\footnote{Discussion of the use of eigenvector decomposition in order to efficiently solve Eq.~\eqref{eq:luscher} in coupled-channel situations can be found in Ref.~\cite{Woss:2020cmp}} of $F^{-1} + \Mc$, where the symmetry of the matrices ensures the orthogonality of the eigenvectors:
\begin{equation*}
 F^{-1} + \Mc  = \sum\nolimits_i \lambda_i \, \mathbf{v}^{}_i\,  \mathbf{v}_i^\intercal,
\end{equation*}
where the eigenvalues and eigenvectors vary with energy.

In order that $\det \big[ F^{-1} + \Mc \big] = 0$ we require at least one eigenvalue to be zero at $E = E_n$. In fact, only a single eigenvalue can be zero, and we label this eigenvalue by $i=0$. In the case that more than one eigenvalue vanished at some energy, the corresponding pole in energy would be of order higher than one and would lead to a correlation function time-dependence incompatible with the time-evolution of discrete energy eigenstates.

Expanding $\lambda_0(E)$ about the zero at $E = E_n$, gives
\begin{equation*}
\lambda_0(E) = (E - E_n)  \left.\frac{d\lambda_0}{dE}\right|_{E_n} + \mathcal{O}\big(E - E_n \big)^2\, ,
\end{equation*}
and from this and Eq.~\eqref{Rn} it is clear that $\widetilde{\mathcal{R}}_n$ is rank-one:
\begin{equation*}
 \widetilde{\mathcal{R}}_n = \frac{2E_n^\star}{{\lambda_0^\star}'}\,  \mathbf{v}^{}_0 \, \mathbf{v}_0^\intercal, 
\end{equation*}
where $\mathbf{v}_0$ is a shorthand for the unit normalized eigenvector evaluated at $E_n$, and where we have used the fact that
\begin{equation*}
2E \cdot \left(\frac{df}{dE} \right)^{-1} = 2E^\star \cdot \left(\frac{df}{dE^\star} \right)^{-1}\, , 
\end{equation*}
and have introduced a $\star'$ notation to indicate differentiation with respect to $E^\star$, 
\begin{equation*}
{\lambda_0^\star}'\equiv \left.\frac{d\lambda_0}{dE^\star}\right|_{E^\star_n} .
\end{equation*}

This eigen-decomposition provides us with the conceptual picture we desired,
\begin{equation*}
\sqrt{2 E_n L^3}\, \big| E_n \big\rangle_L \sim \sqrt{\frac{2 E_n^\star}{ {\lambda_0^\star}'} }\, 
\sum\nolimits_a \big(\mathbf{v}_0\big)_a\,  \big| \varphi_a\varphi_a (E^\star\!=\!E^\star_n)  \big\rangle_\infty \, ,
\end{equation*}
which makes it clear that the finite-volume eigenstates cannot be interpreted as being associated with any one particular hadron-hadron channel, rather they are a linear superpositions of all channels, with weights given by the eigenvector corresponding to the zero eigenvalue. The prefactor which features the slope of the eigenvalue with respect to energy provides the effective  finite-volume normalization of the state, and this can take a value very different to unity. 

In practice it is more convenient to use a slightly different decomposition of $\widetilde{\mathcal{R}}_n$, one which makes use of the eigenvector decomposition of $F + \Mc^{-1}$:
\begin{equation*}
  F + \Mc^{-1}  = \sum\nolimits_i \mu_i \, \mathbf{w}^{}_i \,  \mathbf{w}_i^\intercal \, .
\end{equation*}
Since trivially
\begin{equation*}
F + \Mc^{-1} = F \big( F^{-1} + \Mc \big) \Mc^{-1}\, ,
\end{equation*}
it follows that $\mu_0(E)$ will have a zero at $E = E_n$ just as $\lambda_0(E)$ did, and then since $\big(F + \Mc^{-1}\big)\, \mathbf{w}_0 = 0$ at that energy, we have ${F \mathbf{w}_0 = - \Mc^{-1} \mathbf{w}_0}$. Using the symmetry of the matrices, $\mathbf{w}_0^\intercal F = - \mathbf{w}_0^\intercal \Mc^{-1}$, and we find that Eq.~\eqref{Rn} can be expressed as
\begin{align}
\label{eq:altR}
\widetilde{\mathcal{R}}_n &= 2E_n \cdot \lim_{E \to E_n} (E-E_n) \cdot \, \Mc^{-1} \,  \big( F + \Mc^{-1} \big)^{-1} \, F  \nonumber \\
 &= \left(-\frac{2E^\star_n}{{\mu_0^\star}'}\right) \Mc^{-1} \mathbf{w}^{}_0\, \mathbf{w}_0^\intercal \, \Mc^{-1} \, ,
\end{align}
where all objects are evaluated at $E=E_n$. An example of the numerical determination of $\mu_0$ and $\mathbf{w}_0$ is presented in Appendix~\ref{app:R} where the slope of the zero-crossing eigenvalue is observed to be negative.

A major advantage of the form in Eq.~\eqref{eq:altR} is that it explicitly removes the potentially rapidly energy-varying factor of $\Mc$ from $\Hc$, leaving only the slowly varying $\Ac$,
\begin{equation}
  \label{eq:cdotMap}
 \Big| \big\langle E_n, \mathbf{P}_{\!f}  \big| \mathcal{J}^\mu(0) \big| \textbf{P}_{\!i} \big\rangle_L \Big| = \tfrac{1}{L^3 \sqrt{2 E_i} \sqrt{2 E_n} } \sqrt{-\tfrac{2E^\star_n}{{\mu_0^\star}'}} \; \mathbf{w}_0^\intercal \cdot \left( \tfrac{1}{q^{\star\ell}} \Ac^\mu \right) \, ,
\end{equation}
and this makes clear the importance of the quantities $\mathbf{w}_0$ and $\sqrt{-\tfrac{2 E^\star_n}{{\mu_0^\star}'}}$ for each finite-volume energy level: the first indicates the relative contribution of the various open channels to the finite-volume matrix element, while the second makes the (potentially large) finite-volume correction to the absolute normalization.

\vspace{5mm}
When considering finite-volume energy levels which lie \emph{below} some of the channel thresholds, the matrix $F + \Mc^{-1}$ remains real and symmetric, and the eigenvectors remain orthogonal, so the picture presented above still holds. Elements of $F$ tend to constant values at energies well below a closed threshold in such a way that the closed channel decouples from the scattering system. Details are presented in Appendix~\ref{app:R}.

While at low energies we expect the lowest $\ell$ values to dominate, in general multiple partial-waves will be present, and below kinematic thresholds, we can encounter situations where $F + \Mc^{-1}$ is symmetric, but not purely real. A concrete example might be $\pi K$ scattering in the $A_1$ irrep in moving frames, where both $S$-wave and $P$-wave scattering are present. The symmetry of the matrices ensures that the eigenvectors remain orthogonal ($ \mathbf{w}_i^\intercal \!\cdot\! \mathbf{w}_j = \delta_{ij}$) but in this case, the $P$-wave components of $\mathbf{w}_0$ are now pure imaginary. This does not spoil the reality of the object $\Hc \!\cdot\! \widetilde{\mathcal{R}}_n \!\cdot\! \Hc$ in Eq.~\eqref{eq:LL_formula} owing to the compensating factor $1/q^\star$ present for a $P$-wave in Eq.~\eqref{eq:trans_amp}.

Should a scattering system feature a stable \mbox{bound-state} in a particular partial-wave, lying well below all kinematic thresholds, the properties of $F$ are such that Eq.~\eqref{eq:LL_formula} reduces to the volume-independent result we would expect for a transition between stable hadrons. More discussion is presented in Appendix~\ref{app:R}.~\footnote{For a recent discussion of this scenario in the context of $2 \xrightarrow{\mathcal{J}} 2$ matrix elements see Ref.~\cite{Briceno:2019nns}.}

\vspace{5mm}

In the remainder of this paper we will present examples of the implementation of the approach presented in this section. The idea is that the scattering matrix for some number of coupled hadron-hadron channels is determined using energy-dependent parameterizations of $\Mc$ to describe finite-volume spectra, along the lines described in detail in Ref.~\cite{Briceno:2017max}. The eigen-decomposition of $F + \Mc^{-1}$ can then be carried out for the parameterized $\Mc$ at each solution of $\det \big[ F + \Mc^{-1} \big] = 0$ corresponding to a calculated finite-volume energy level. By evaluating the eigenvalues in the neighborhood of the finite-volume energy, one can compute the derivative of the eigenvalue and hence an implementation of Eq.~\eqref{eq:cdotMap} can be set up for each calculated finite-volume matrix element. 
These are linear equations featuring unknowns, $\Ac_{a, \ell m_\ell}(Q^2,s)$, as well as the known finite-volume matrix elements (coming from explicit lattice QCD computation of three-point correlation functions). By parameterizing the $Q^2$-- and $s$--dependence, we can minimize a $\chi^2$ built out of all the implementations of Eq.~\eqref{eq:cdotMap} for the various kinematic points computed.

We will illustrate the approach using some simple toy-models of coupled hadron-hadron scattering showing that with a modest number of computed finite-volume transition matrix elements, the corresponding infinite-volume result can be determined. Having constrained such transition amplitudes in cases where the scattering system features a resonance, we will also demonstrate that the continuation to the resonant pole can be performed leading to the transition form-factor of the resonance.

%%%%%%%%%%%%%%%%%%%%%%%%%%%%%%%%%%%%%%%%
\pagebreak %%%%%%%%%%%%%%%%%%%%%%%%%%%%%
%%%%%%%%%%%%%%%%%%%%%%%%%%%%%%%%%%%%%%%%

%%%%%%%%%%%%%%%%%%%%%%%%%%%%%%%%%%%%
% Section III - the toy models
%%%%%%%%%%%%%%%%%%%%%%%%%%%%%%%%%%%%
\section{Toy models of scattering and transitions in infinite and finite volume}\label{sec:III}
%

%%%%%%%%%%%%%%%%%%%%%%%%%%%%%%%%%%%%%%%%%%%%%%%%%%%%%%%%%%%%%%%%%%
\begin{figure*}[t]
  \begin{center}
\includegraphics[width=0.9\textwidth]{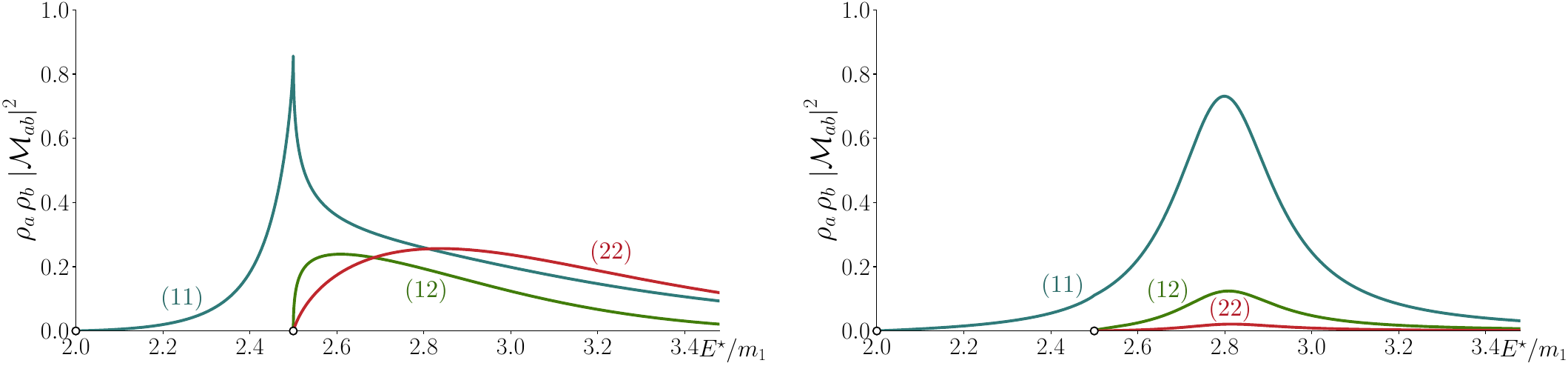}
  \caption{(a) ``\emph{cusp-like}'' and (b) ``\emph{Flatt\'e-like}'' scattering matrices shown as a function of $E^\star$ via $\rho_1^2 \, |\mathcal{M}_{11}|^2$ (blue), $\rho_1\rho_2 \, |\mathcal{M}_{12}|^2$ (green), and $\rho^2_2 \, |\mathcal{M}_{22}|^2$ (red). Open circles on the axis indicate kinematic thresholds for channels 1 and 2. 
  }
  \label{fig:cusp_flatte_amps}
  \end{center}
\end{figure*}
%%%%%%%%%%%%%%%%%%%%%%%%%%%%%%%%%%%%%%%%%%%%%%%%%%%%%%%%%%%%%%%%%%

%%%%%%%%%%%%%%%%%%%%%%%%%%%%%%%%%%%%%%%%%%%%%%%%%%%%%%%%%%%%%%%%%%
\begin{figure*}[t!]
  \begin{center}
\includegraphics[width=0.9\textwidth]{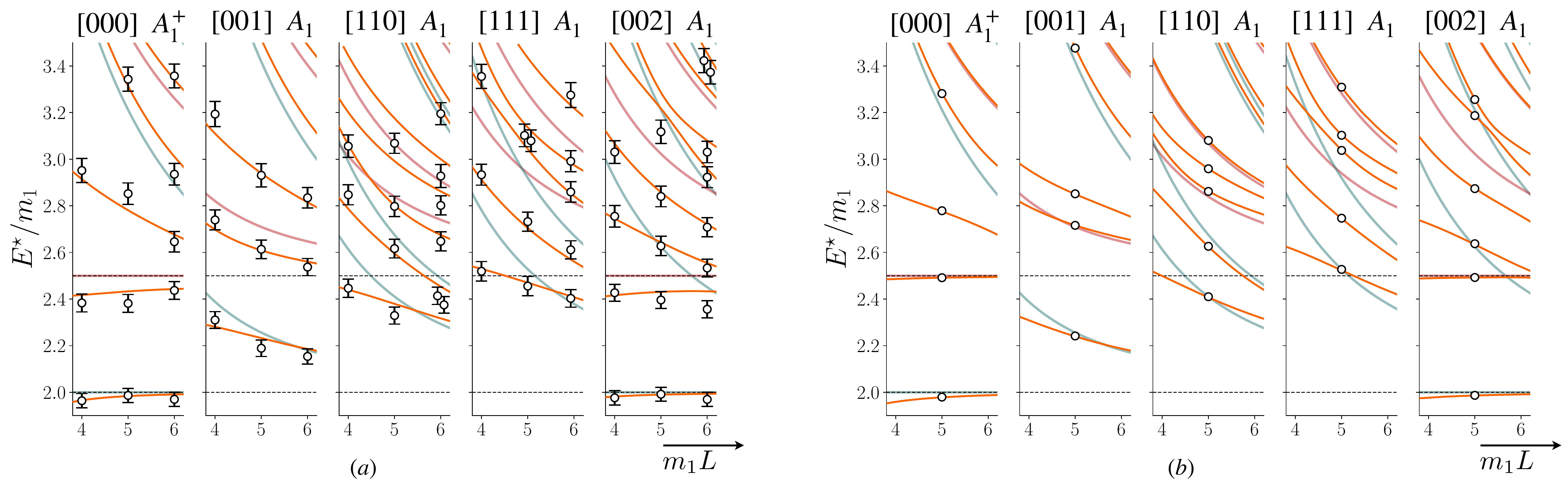}
    \caption{ Orange curves indicate the finite volume spectra obtained by solving Eq.~\ref{eq:luscher} for the (a) ``\emph{cusp-like}'' and (b) ``\emph{Flatt\'e-like}'' amplitudes described in the text. Each panel shows the spectrum for a different total momentum $\textbf{P}=\tfrac{2\pi}{L}\,\big[{n}_x {n}_y {n}_z \big]$. Blue and red lines show the spectrum of $\varphi_1 \varphi_1$ and $\varphi_2 \varphi_2$ states that would be present in a non-interacting theory. 
	In (a) the points with errorbars show a synthetic spectrum on three volumes generated with statistical uncertainties, to be described in Section~\ref{sec:IV}. In (b) the points without errorbars show the exact spectrum on a single volume to be considered in Section~\ref{sec:IV}.
   }
  \label{fig:spectra}
  \end{center}
\end{figure*}
%%%%%%%%%%%%%%%%%%%%%%%%%%%%%%%%%%%%%%%%%%%%%%%%%%%%%%%%%%%%%%%%%%

We choose to model a system of two coupled hadron-hadron channels, $(\varphi_1 \varphi_1, \varphi_2 \varphi_2)$, where $\varphi_1$, $\varphi_2$ are scalar mesons with masses $m_1, m_2= 1.25\, m_1$ respectively. These channels can be reached from a production process $\gamma \chi$ where $\chi$ is a scalar meson of mass $M = 1.25\, m_1$, and where $\gamma$ represents the action of a conserved vector current. We will examine two simple toy-models describing coupled-channel resonant scattering in $S$-wave and the corresponding transition amplitudes describing $\gamma \chi \to \varphi_1 \varphi_1$ and $\gamma \chi \to \varphi_2 \varphi_2$.

%%%%%%%%%%%%%%%%%%%%%%%%%%%%%%%%%%%%%%%%%%%%%%%%%%%%%%%%%
\subsection{Toy model scattering amplitudes}

We can ensure coupled-channel unitarity, described by Eq.~\eqref{eq:unitarity}, by making use of a $K$-matrix parameterization in
\begin{equation*}
\mathcal{M}^{-1} = \mathcal{K}^{-1} - i \rho \,,
\end{equation*}
where $\mathcal{K}$ is a symmetric matrix taking real values for all real energies. We will explore two models in which the $K$-matrix has elements
\begin{equation} \label{eq:K}
\mathcal{K}_{ab}(s) = \frac{g_a \, g_b}{m^2 -s} + \gamma_{ab}\, ,
\end{equation}
where constants $m, g_1, g_2, \gamma_{11}, \gamma_{12}, \gamma_{22}$ are parameters. We will see that two different choices of values for these lead to qualitatively rather different amplitudes.

%%%%%%%%%%%%%%%%%%%%%%%%%%%%%%%%%%%%%%%%%%%%%%%%%%%%%%%%%%
%% cusp-like
\vspace{3mm}
First we construct a resonant amplitude with very strong coupling between the channels, featuring a \mbox{``\emph{cusp-like}''} enhancement at the second threshold. The parameters values, $\big\{$$m/m_1 \!=\! 2.60$, $g_1/m_1 \!=\! 15$, $g_2/m_1 \!=\! 19$, $\gamma_{11} \!=\! 0.1$, $\gamma_{12} \!=\! 30$, $\gamma_{22} \!=\! 11$$\big\}$, generate this amplitude, where the elements of the resulting $\mathcal{M}$ are shown for real scattering energies in Figure~\ref{fig:cusp_flatte_amps}(a). 

While a cusp in $\varphi_1 \varphi_1 \to \varphi_1 \varphi_1$ is generically expected at the opening of the $\varphi_2 \varphi_2$ channel, due to the square-root in $\rho_2(s)$, the strength of the observed feature, and the rapid turn-on of amplitudes leading to the $\varphi_2 \varphi_2$ final-state suggests resonant behaviour, and indeed this amplitude is found to have a nearby pole singularity at 
${\sqrt{s_R}/m_1 = 2.59-\frac{i}{2}\,0.13}$ on unphysical sheet {\sf II}\footnote{For a two channel scattering system, there are three unphysical Riemann sheets. Above the first threshold, but below the second, sheet {\sf II} is closest to physical scattering, while above both thresholds, the proximal sheet is {\sf III}. More detailed discussion of sheet structure can be found in Ref.~\cite{Briceno:2017max} and references therein.
}.
This pole has couplings $c_1, c_2$ which have comparable magnitudes, indicating that this resonance couples strongly to both channels.

%%%%%%%%%%%%%%%%%%%%%%%%%%%%%%%%%%%%%%%%%%%%%%%%%%%%%%%%%%
%% flatte-like
\vspace{3mm}
Our second amplitude choice, which we will describe as ``\emph{Flatt\'e-like}'', reflects a more straightforward picture of a coupled-channel resonance. Using the parameter values, 
$\big\{$$m/m_1 = 2.80$,   $g_1/m_1 = 10$, $g_2/m_1 = 5$, ${\gamma_{11} = 0.01}$, $\gamma_{12} = 0$, $\gamma_{22} = 0.5$$\big\}$, we obtain the $\mathcal{M}$ elements shown in Figure~\ref{fig:cusp_flatte_amps}(b), which we observe to be simply an isolated ``bump'' lying above both thresholds. Examined for complex values of the energy, the bump reflects the presence of a sheet {\sf III} pole lying very close to the real energy axis at $\sqrt{s_R}/m_1 = 2.80 - \frac{i}{2}0.21$
\footnote{An additional ``mirror'' pole, less relevant to the bump region by virtue of being more distant, is present on sheet {\sf II}. }
. The pole couplings $c_1, c_2$, which are close to being real, have magnitudes which closely reflect the hierarchy selected for the $K$-matrix parameters $g_1, g_2$, indicating a significantly weaker coupling of the resonance to $\varphi_2 \varphi_2$ relative to $\varphi_1 \varphi_1$.

This second amplitude is ``\emph{Flatt\'e-like}'' in that were it not for the small non-zero values of $\gamma_{11}, \gamma_{22}$, it would be of the form,
\begin{align*}
\mathcal{M}^\mathrm{Fl.}_{ab} &= \frac{g_a \, g_b}{D(s)} \, ,\\
 &\mathrm{where} \; D(s) = m^2 -s - i g_1^2 \,\rho_1(s) - i g_2^2 \,\rho_2(s) \, ,
\end{align*}
commonly referred to as the Flatt\'e-form~\cite{Flatte:1976xu}. Such an amplitude intuitively describes a single resonance coupled to two channels with no ``background'', but its simplicity gives rise to some rather peculiar properties. These follow from the fact that the scattering matrix \emph{factorizes}, such that even in an $N$-channel case where ${\mathcal{M} = \mathbf{g} \, \mathbf{g}^\intercal \, D^{-1}(s) }$ with ${\mathbf{g}= (g_1, g_2 \ldots g_N)}$, the matrix $\mathcal{M}$ has a rank of only $1$, having one non-zero eigenvalue with eigenvector $\mathbf{g}$, and $N-1$ zero eigenvalues with eigenvectors orthogonal to $\mathbf{g}$
\footnote{One immediate consequence of this is that $\mathcal{M}^{-1}$ does not exist at any energy for the Flatt\'e amplitude. The small non-zero $\gamma$ values in our amplitude choice regulate the singular nature such that $\mathcal{M}^{-1}$ does exist.}.
% 

%%%%%%%%%%%%%%%%%%%%%%%%%%%%%%%%%%%%%%%%%%%%%%%%%%%%%
\vspace{3mm}
Given an amplitude parameterization, and a set of parameter values, we can solve the quantization condition, Eq.~\eqref{eq:luscher}, in several volumes and moving frames\footnote{The subduction into $[000]A_1^+$ and moving frame $A_1$ irreps is trivial for these purely $S$-wave amplitudes.}.
The volumes and frames selected are designed to mimic accessible cases considered in contemporary lattice QCD calculations~\cite{Dudek:2014qha, Wilson:2014cna, Briceno:2017qmb, Dudek:2016cru, Woss:2019hse, Moir:2016srx, Woss:2020ayi}. For the ``\emph{cusp-like}'' and ``\emph{Flatt\'e-like}'' amplitudes, the resulting finite-volume spectra are shown in Figure~\ref{fig:spectra}.

%%%%%%%%%%%%%%%%%%%%%%%%%%%%%%%%%%%%%%%%%%%%%%%%%%%%%
%%%%%%%%%%%%%%%%%%%%%%%%%%%%%%%%%%%%%%%%%%%%%%%%%%%%%
%%%%%%%%%%%%%%%%%%%%%%%%%%%%%%%%%%%%%%%%%%%%%%%%%%%%%
\subsection{Toy model transition amplitudes and finite-volume matrix elements}

Equation~\eqref{eq:HPFM} relates the transition amplitudes $\mathcal{H}_a$ to the scattering matrix $\mathcal{M}$ and Lorentz-invariant transition form-factors $\mathcal{F}_a(Q^2, s)$. In order to proceed further with our toy-modelling exercise, we must make explicit choices for the current-virtuality ($Q^2$) and scattering energy ($s$) dependence of the form-factors.

In principle, the form-factors are subject to constraints, for example those arising from unitarity applied to the  crossed-channels. In this first investigation we will not attempt to implement these constraints, which primarily impact significantly time-like (negative) values of $Q^2$, instead focussing on values of $Q^2$ in or close to the \mbox{space-like} region. Our intention is to test the practicality of extracting the form-factors from finite-volume matrix-elements, and for this exercise their detailed analytical structure in $Q^2$ is not of primary interest.

With this discussion in mind, we can construct a range of parametrizations for the form factors, similar to the ones used to describe $\gamma \pi \to \pi \pi$ in Refs.~\cite{Briceno:2016kkp, Briceno:2015dca}. We discuss this broad class of parametrization in Sec.~\ref{sec:fixM_varyF}, where they will be used in fits. Here we select one parameterization, which will serve as our underlying model from which we generate synthetic data: 
\begin{equation}
\label{eq:Ftoy}
\widetilde{\mathcal{F}}_a(Q^2, s) = m_1 \, \frac{ f_a^{(0)} + f_a^{(1)} \tfrac{s}{m_1^2}  }{ Q^2 + m_Q^2} \, ,
\end{equation}
where the $f_a^{(i)}$ coefficients are dimensionless. 
As previously mentioned, $s$-channel unitarity applied to the transition amplitudes, $\Hc_a(s)$, ensures that the form-factors do not have any singularities in $s$, justifying a polynomial-in-$s$ construction. The choice of a simple-pole in $Q^2$ generates a typical monotonically decreasing behavior in the \mbox{space-like} region. It is worth noting that this parametrization does respect one analytic property of the form-factors, namely that any singularities in $Q^2$ be independent of the hadron-hadron channel produced -- this is manifest in the fact that the pole location, $m_Q^2$, does not carry a channel index. 

\begin{figure*}[t]
  \begin{center}
\includegraphics[width=0.8\textwidth]{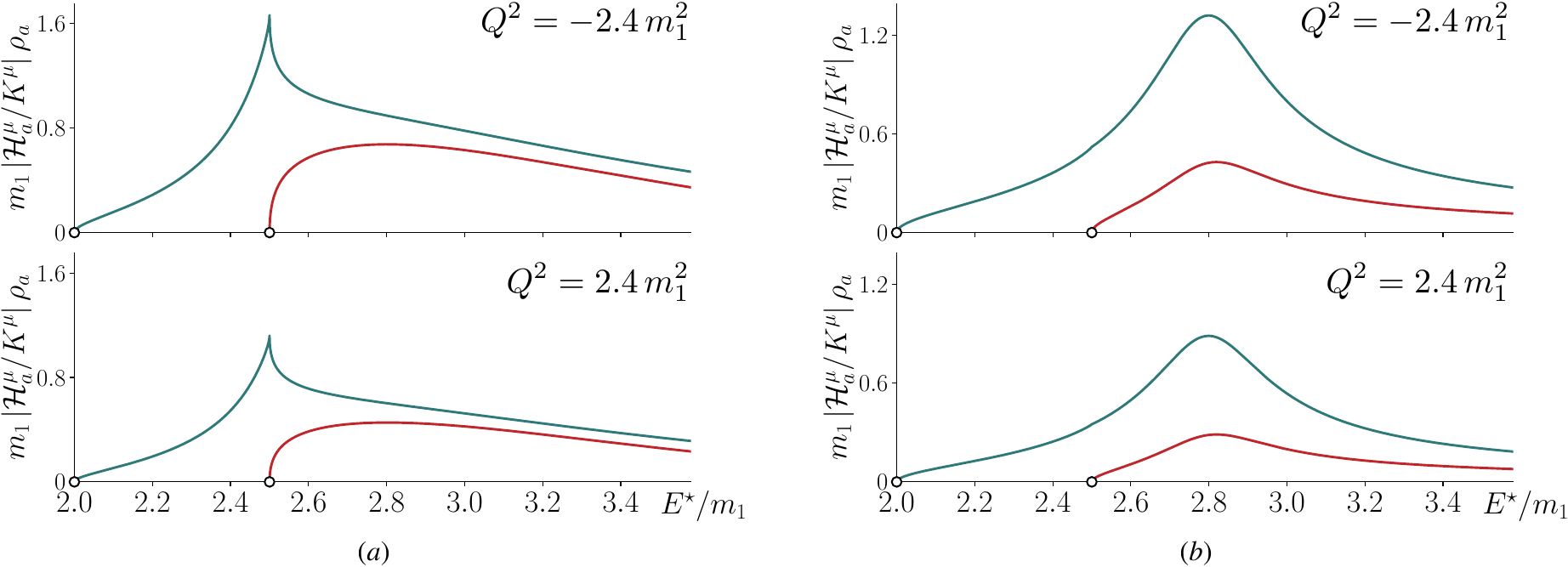}
  \caption{ Transition matrix elements for (a) ``\emph{cusp-like}'' and (b) ``\emph{Flatt\'e-like}'' amplitudes using the form-factor model in Eq.~\eqref{eq:Ftoy}. Plotted is the ratio $|\mathcal{H}^\mu/K^\mu|\, \rho$, which removes the trivial kinematic factor, for production of channel 1 (blue) and channel 2 (red). The amplitudes are plotted as functions of the final-state energy for two values of the current virtuality, $Q^2$.}
  \label{fig:Hplots}
  \end{center}
\end{figure*}

We select parameter values, $\big\{$$m_Q= 3.5\, m_1$, ${f_1^{(0)} = 12.25}$, $f_1^{(1)}  = 0.1$, $f_2^{(0)} = 1$, $f_2^{(1)} = 0.4$$\big\}$, and with our model choices for $\mathcal{F}$ and $\mathcal{M}$ in hand, we can construct transition amplitudes according to Eq.~\eqref{eq:HPFM}. Given that the overall kinematic pre-factor is in general non-zero and finite, we can divide the transition amplitude by this factor. For convenience, we will define, 
\begin{align}
K^\mu \equiv  (P_i+ P_f)^\mu  + \tfrac{s-M^2}{Q^2} (P_f - P_i)^\mu \, , 
\label{eq:Kmu}
\end{align}
and using this the dynamical quantity that one hopes to constrain is the scalar ratio $\Hc^\mu/K^\mu$.  The energy-dependence of these amplitudes at a sample set of $Q^2$ values is shown in Figure~\ref{fig:Hplots}.

%%%%%%%%%%%%%%%%%%%%%%%%%%%%%%%%%%%%%%%%%%%%%%%%%%%%%%%%%%%%%%%%%%
%%%%%%%%%%%%%%%%%%%%%%%%%%%%%%%%%%%%%%%%%%%%%%%%%%%%%%%%%%%%%%%%%%

%% flatte form
\vspace{3mm}
Note that the similarity of our ``\emph{Flatt\'e-like}'' amplitude, $\Mc$, to a factorizing rank-one form has an important impact on the properties of the transition amplitude. For the Flatt\'e amplitude, the combination
\begin{equation*}
 \sum\nolimits_b \mathcal{F}_b \, \mathcal{M}^{\mathrm{Fl.}}_{ba} = \big( \bm{\mathcal{F}}^\intercal \!\cdot\! \mathbf{g} \big) 
 \, g_a \, D^{-1}(s) \, ,
\end{equation*}
which in the two-channel case depends upon $\mathcal{F}_1, \mathcal{F}_2$ only in the combination ${g_1 \mathcal{F}_1 + g_2 \mathcal{F}_2}$, and as such one should not expect to be able to determine $\mathcal{F}_1, \mathcal{F}_2$ separately 
\footnote{In this two-channel case, there is a single orthogonal combination $g_2 \mathcal{F}_1 - g_1 \mathcal{F}_2$ which is inaccessible.}
. The intuitive origin of this effect is that rather than the general case of $\gamma \chi$ producing either $\varphi_1 \varphi_1$ or $\varphi_2 \varphi_2$ which then rescatter into each other, for the Flatt\'e amplitude, in effect $\gamma \chi$ at any real energy can only produce ``the resonance'' (in this case meaning the energy dependence $D^{-1}(s)$ rather than the complex-$s$ pole). This suggests an interpretation of $g_1 \mathcal{F}_1 + g_2 \mathcal{F}_2$ as an effective coupling $g( \gamma \chi \to \text{`\!}R\text{'})$. In a more general amplitude, this exact factorization would hold \emph{only} at the complex resonance pole position, $s = s_R$, and in principal it should be possible to determine $\mathcal{F}_1, \mathcal{F}_2$ separately for real energies for generic amplitudes.

%%%%%%%%%%%%%%%%%%%%%%%%%%%%%%%%%%%%%%%%%%%%%%%%%%%%%
%% matrix elements in a finite volume
\vspace{3mm}

Our approach to generating synthetic finite-volume matrix-element data is to make use of Eq.~\eqref{eq:cdotMap}. For our toy-model amplitudes we can find the values of ${\mu_0^\star}'$ and $\mathbf{w}_0$ for each finite-volume energy-level shown in Figure~\ref{fig:spectra}. Given the parameterizations of $\mathcal{F}_{1,2}$ described above, we can hence construct matrix-element values for a number of discrete kinematical points where the initial $\chi$ state has an allowed lattice momentum,
\begin{equation*}
P_i^\mu = \Big( \sqrt{M^2 + \left(\tfrac{2\pi}{L}\right)^2 |\mathbf{n} |^2 }, \tfrac{2\pi}{L} \mathbf{n} \Big)\, ,
\end{equation*}
and where the final-state is one of the finite-volume energy eigenstates from Figure~\ref{fig:spectra}.

\begin{figure}[h!]
\includegraphics[width=.5\textwidth]{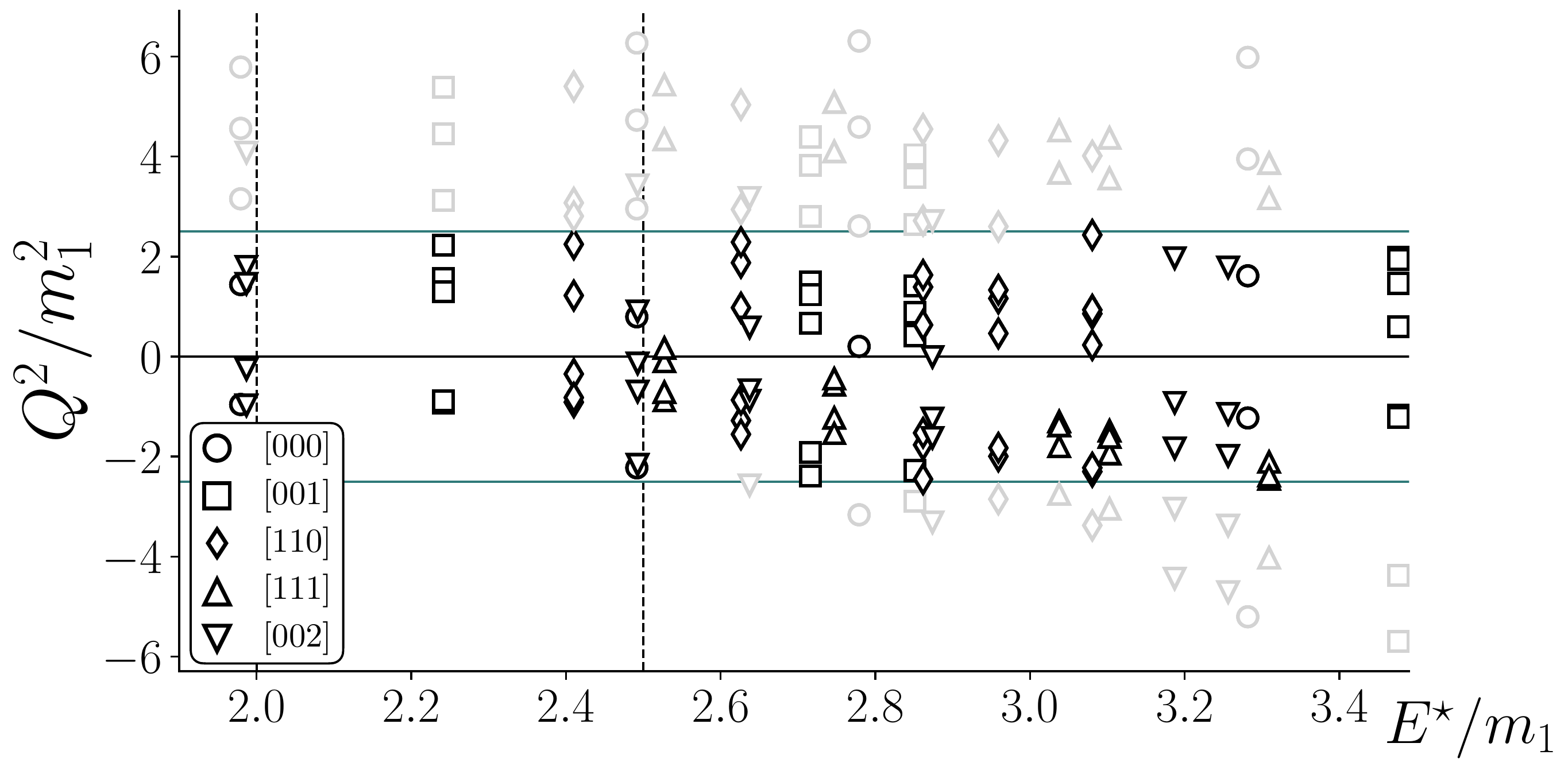}
  \caption{ The distribution of kinematically accessible points for the ``\emph{Flatt\'e-like}'' amplitude for a single volume of $L=5/m_1$.  The legend labels the total momentum of the final hadron-hadron state in units of $2\pi/L$. The channel thresholds are depicted by the vertical dashed lines. The black points are retained after application of kinematic cuts as described in the text. }
  \label{fig:Q2vsE}
\end{figure}

In generating the possible kinematic points, we limit the spatial momentum of the initial/final states and the current to have $|\textbf{n}|^2 \leq 4$, and we restrict to a region of virtuality $-2.5\, m_1^2 < Q^2 < 2.5\, m_1^2$, and final-state energy $E^\star < 3.5\, m_1$. The resulting set of points is comparable to those which can be obtained in contemporary lattice QCD calculations, while still giving a broad range of kinematic constraint on the desired transition amplitudes. Figure~\ref{fig:Q2vsE} illustrates the kinematic coverage for the \mbox{``\emph{Flatt\'e-like}''} amplitude using only a single volume of $L=5/m_1$ -- the \mbox{``\emph{cusp-like}''} amplitude has a comparable coverage.

%%%%%%%%%%%%%%%%%%%%%%%%%%%%%%%%%%%%%%%%%%%%%%%%%%%%%%%%%%%%%%%%%%
%%%%%%%%%%%%%%%%%%%%%%%%%%%%%%%%%%%%%%%%%%%%%%%%%%%%%%%%%%%%%%%%%%

\vspace{5mm}

In performing a global analysis of matrix elements over a range of kinematic points, it is convenient to extract the trivial kinematic factor present in Eq.~\eqref{eq:Amua}. We achieve this by defining
\begin{equation}
  \label{eq:FL}
\mathcal{F}_L(Q^2, s) \equiv  \sqrt{-\frac{2E^\star_n}{{\mu_0^\star}'}} \; \mathbf{w}_0^\intercal \!\cdot\! \mathcal{F}(Q^2, s)  \, ,
\end{equation} 
and using this Eq.~\eqref{eq:cdotMap} becomes
\begin{equation*}
 \Big| \big\langle E_n, \mathbf{P}_{\!f}  \big| \mathcal{J}^\mu(0) \big| \textbf{P}_{\!i} \big\rangle_L \Big| = \tfrac{1}{L^3 \sqrt{2 E_i} \sqrt{2 E_n} }\, K^\mu \, \Fc_L\, ,
\end{equation*}
where the normalizations are such that in the limit of an infinitesimally narrow resonance, $\mathcal{F}_L(Q^2, s)$ coincides with the definition of the $1 \xrightarrow{\mathcal{J}} R$ form-factor, treating $R$ as a stable particle~\footnote{For the electromagnetic current, $\Fc_L$ is dimensionless, as expected for transition form factors coupling scalar initial/final states.}.

%%%%%%%%%%%%%%%%%%%%%%%%%%%%%%%%%%%%%%%%%%%%%%%%%%%%%%%%%%%%%%%%%%
\begin{figure*}[t]
  \begin{center}
\includegraphics[width=.84\textwidth]{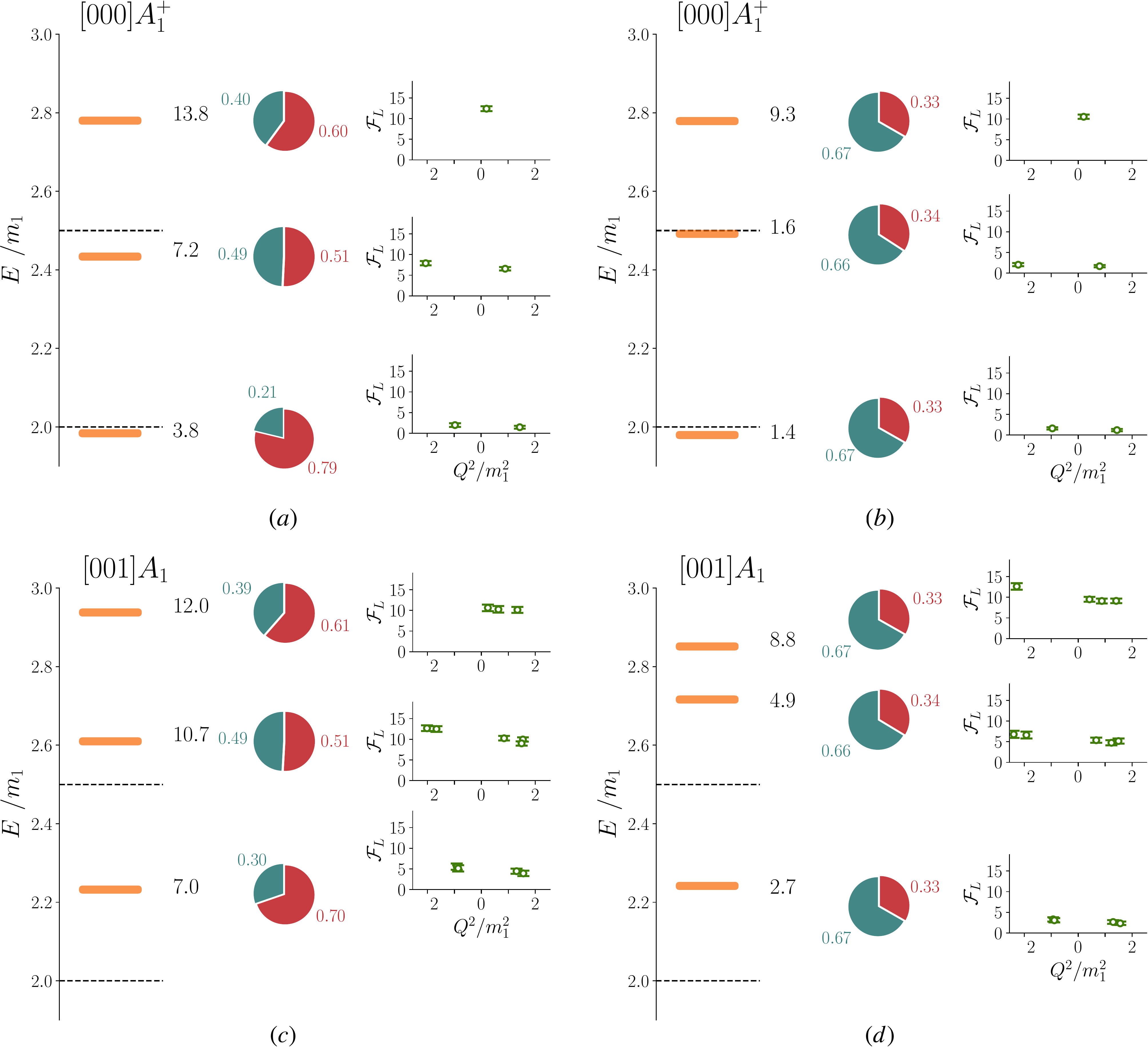}
  \caption{ Finite volume spectrum for $L =5/m_1$ in two frames for (a,c) ``\emph{cusp-like}'' and (b,d) ``\emph{Flatt\'e-like}'' amplitudes. Shown alongside each energy level are the corresponding values of $\sqrt{\frac{-2 E^\star_n}{{\mu_0^\star}'}}$ (black numbers, in units of $m_1$),  $\mathbf{w}_0$ (pie chart, blue for channel 1, red for channel 2), and $\mathcal{F}_L$ (as defined in Eq.~\eqref{eq:FL}, green points). The pie charts shown indicate $\tfrac{({\bf w_0})_i}{({\bf w_0})_1 + ({\bf w_0})_2}$, and the discrete $Q^2$ values correspond to the set of kinematic points previously plotted in Figure~\ref{fig:Q2vsE}. The origin of the errorbars shown for the $\mathcal{F}_L$ points will be described in Section~\ref{sec:IV}. 
  } \label{fig:pie_charts}
  \end{center}
\end{figure*}
%%%%%%%%%%%%%%%%%%%%%%%%%%%%%%%%%%%%%%%%%%%%%%%%%%%%%%%%%%%%%%%%%%

Figure~\ref{fig:pie_charts} illustrates the behavior of ${\mu_0^\star}'$ and $\mathbf{w}_0$ for the ``\emph{Flatt\'e-like}'' and ``\emph{cusp-like}'' models in a volume ${L=5/m_1}$ for the rest frame and one moving frame. For each \mbox{finite-volume} energy level, we provide the value of $\sqrt{-\frac{2 E^\star_n}{{\mu_0^\star}'}}$ and the relative sizes of $(\mathbf{w}_0)_1$ and $(\mathbf{w}_0)_2$ \footnote{The figure shows pie charts constructed as $\tfrac{({\bf w}_0)_i}{({\bf w}_0)_1 + ({\bf w}_0)_2}$ to show the relative sizes of $({\bf w}_0)_1$, $({\bf w}_0)_2$, but in Eq.~\eqref{eq:FL}, the unit-normalized vector (${ {\bf w}_0^\intercal \!\cdot\! {\bf w}_0 = 1 }$) should be used.}. These illustrate two important features of the coupled-channel finite-volume formalism: 
First, that the magnitude of the finite-volume scaling characterized by $\sqrt{-\frac{2 E^\star_n}{{\mu_0^\star}'}}$ is in general \emph{not close to unity}, and varies significantly \mbox{level-by-level} -- this is just one indication of serious systematic errors that can be introduced if matrix-elements of unstable hadrons are studied ignoring effects due to the finite-volume of the lattice. 
Second, we see a very different behavior for $\mathbf{w}_0$ depending on the model. For the ``\emph{cusp-like}'' model the relative sizes of the components of the eigenvector, $(\mathbf{w}_0)_1, (\mathbf{w}_0)_2$, change depending upon the energy level, indicating that each discrete finite-volume energy is sensitive to a different linear combination of $\mathcal{F}_1, \mathcal{F}_2$ evaluated at that energy.  On the other hand, for the ``\emph{Flatt\'e-like}'' amplitude, the ratio $\tfrac{(\mathbf{w}_0)_2 }{ (\mathbf{w}_0)_1 }$ appears to take essentially the same value for every energy level, one which is extremely close to the \mbox{value of $\tfrac{g_2}{g_1}$}. This was to be expected, and simply reflects the impact of the near-rank-one nature of the ``\emph{Flatt\'e-like}'' amplitude on the finite-volume spectrum. In Appendix~\ref{app:Flatte} we show that an \mbox{$N$-channel} Flatt\'e amplitude will always have $\mathbf{w}_0 \propto \mathbf{g}$ for every finite-volume energy eigenstate. Our addition of small values for $\gamma_{11}, \gamma_{22}$ has a negligible impact on $\mathbf{w}_0$ in the energy region we consider. A consequence of this is to ensure that our previous expectation, that only the combination $g_1 \mathcal{F}_1 + g_2 \mathcal{F}_2$ should be accessible, remains true in a finite-volume.

\pagebreak
%%%%%%%%%%%%%%%%%%%%%%%%%%%%%%%%%%%%
% Section IV - fake data analysis
%%%%%%%%%%%%%%%%%%%%%%%%%%%%%%%%%%%%
\section{Extraction of infinite-volume transition matrix-elements from synthetic finite-volume data}\label{sec:IV}

With our toy models defined, we may generate synthetic data with errors designed to resemble that which can be obtained in contemporary lattice QCD calculations, and with that data in hand we can attempt to reconstruct the $s$ and $Q^2$ dependence of the input transition amplitudes using parameterizations. We will consider two situations: the first assumes (unrealistically) that we know the \emph{exact} scattering amplitude, such that only the form-factors need to be parameterized, while in the second, more realistic case, we must also determine $\Mc(s)$ using finite-volume spectrum data with errors.

\pagebreak
%%%%%%%%%%%%%%%%%%%%%%%%%%%%%%%%%
\subsection{Idealized situation}\label{sec:fixM_varyF}
%%%%%%%%%%%%%%%%%%%%%%%%%%%%%%%%%

%%%%%%%%%%%%%%%%%%%%%%%%%%%%%%%%%%%%%%%%%%%%%%%%%%%%%%%%%%%%%%%%%%
\begin{figure*}
  \begin{center}
\includegraphics[width=0.72\textwidth]{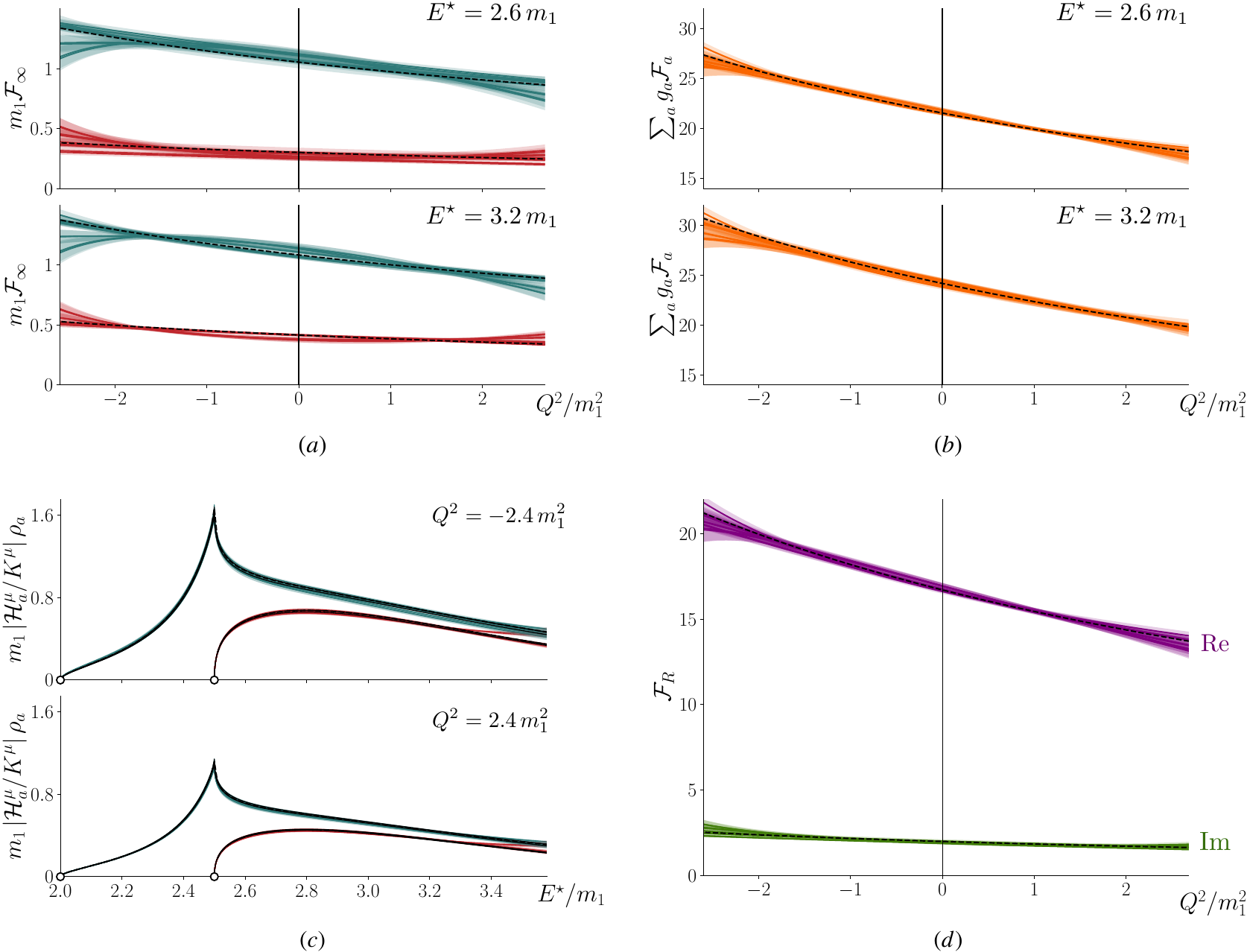}
  \caption{For the ``\emph{cusp-like}'' amplitude, 11 successful descriptions of the $\Fc_L$ synthetic data using the exact $\Mc$. In all panels the dashed black line shows the exact input function. (a) The infinite-volume form-factors, $\Fc_1$ (blue), $\Fc_2$ (red), at two values of $E^\star$. (b) Sums of $\Fc_a$ weighted by the channel couplings $g_a$ used in Eq.~\eqref{eq:K}. (c) The transition amplitudes. (d) The \emph{transition form-factor of the resonance} obtained from the residue at the pole, real part in purple, imaginary part in green.}
  \label{fig:Cusp_fits}
  \end{center}
\end{figure*}
%%%%%%%%%%%%%%%%%%%%%%%%%%%%%%%%%%%%%%%%%%%%%%%%%%%%%%%%%%%%%%%%%%

We explore the sensitivity of the finite-volume process to the transition form-factors by first assuming that the scattering matrix $\Mc(s)$ is known exactly, but that we still have only a limited number of finite-volume matrix elements evaluated at final-state energies corresponding to the discrete spectrum in a finite-volume. We choose to use only those kinematic points presented in Figure~\ref{fig:Q2vsE}, and generate synthetic data for $\Fc_L$ defined in Eq.~\eqref{eq:FL} by drawing from a gaussian probability distribution such that the points have a 5\% error and a random fluctuation of the mean value commensurate with that. For simplicity, the values at different kinematics are treated as being uncorrelated. 

This procedure applied to the ``\emph{cusp-like}'' amplitude leads to values of $\Fc_L$ at 80 kinematic points across 22 $E^\star_n$ levels, and for the ``\emph{Flatt\'e-like}'' amplitude, 92 values across 24 $E^\star_n$ levels. For comparison relatively recent lattice QCD calculations of \emph{elastic} $\pi \gamma \to \pi \pi$~\cite{Briceno:2015dca, Alexandrou:2018jbt}, each on a single volume, featured 42 and 48 kinematic points respectively. Considering that a coupled-channel system inevitably leads to an increase in the density of \mbox{finite-volume} energy levels and hence the number of accessible kinematic points, our data set appears to be a quite reasonable estimate of the number of points that will be available in forthcoming calculations.

Given these $\Fc_L$ data, we minimize a $\chi^2$ assuming a parameterization for the channel form-factors $\Fc_{a=1,2}(Q^2,s)$, treating this form as essentially unknown. As such we explore a range of possible parameterizations. In all cases, a low-order polynomial in $s$ is considered, consistent with a lack of $s$-channel singularities (with one caveat, see below). In $Q^2$, polynomials, a simple-pole and exponential forms are used. The following form captures the variations,
\begin{equation}
\Fc_a(Q^2, s) = \frac{ b_a^{(0)} + b_a^{(1)}\!\cdot\!s + b_a^{(2)}\!\cdot\!s^2 \; +\; c_a^{(1)}\!\cdot Q^2 + c_a^{(2)}\!\cdot Q^4  }{ \alpha \big( m_P^2 + Q^2 + d_a\!\cdot\! s \big) + \beta e^{ Q^2/r^2 } + \gamma     } \, ,
\end{equation}
by selecting $\alpha, \beta, \gamma$ to be 1 or 0. The $d_a\!\cdot\! s$ term in the denominator does allow for ($Q^2$-dependent) poles in $s$, which can be viewed as a very crude approximation to having a left-hand cut in the transition amplitude.

A second class of parameterization, which might be argued to be less model-dependent, makes use of a mapping to a variable $z$ for which a polynomial form is expected to converge rapidly~\cite{Boyd:1994tt,Boyd:1997qw,Bourrely:2008za,Bourrely:2008za}. $z$ is defined as,
\begin{equation*}
z(Q^2) = \frac{\sqrt{t_c +Q^2} - \sqrt{t_c} }{\sqrt{t_c + Q^2} + \sqrt{t_c}}\, ,
\end{equation*}
where $t_c = (2m_1)^2$ is the position of the nearest hypothetical branch-point singularity in $-Q^2$. The coefficients in the polynomial of $z$ are allowed to be \mbox{low-order polynomials in $s$},
\begin{align*}
\Fc_a(Q^2, s) &= \sum_{n=0} a_n(s) \, z^n \nonumber \\
&\mathrm{where}\;\; a_n(s) = \sum_{m=0} a^{(n)}_m s^m \, ,
\end{align*}
where in practice we allow up to quadratic order in each of $z$, $s$.

%%%%%%%%%%%%%%%%%%%%%%%%%%%%%%%%%%%%%%%%%%%%%%%%%%%%%%%%%%%%%%%%%%
\begin{figure*}
  \begin{center}
\includegraphics[width=0.72\textwidth]{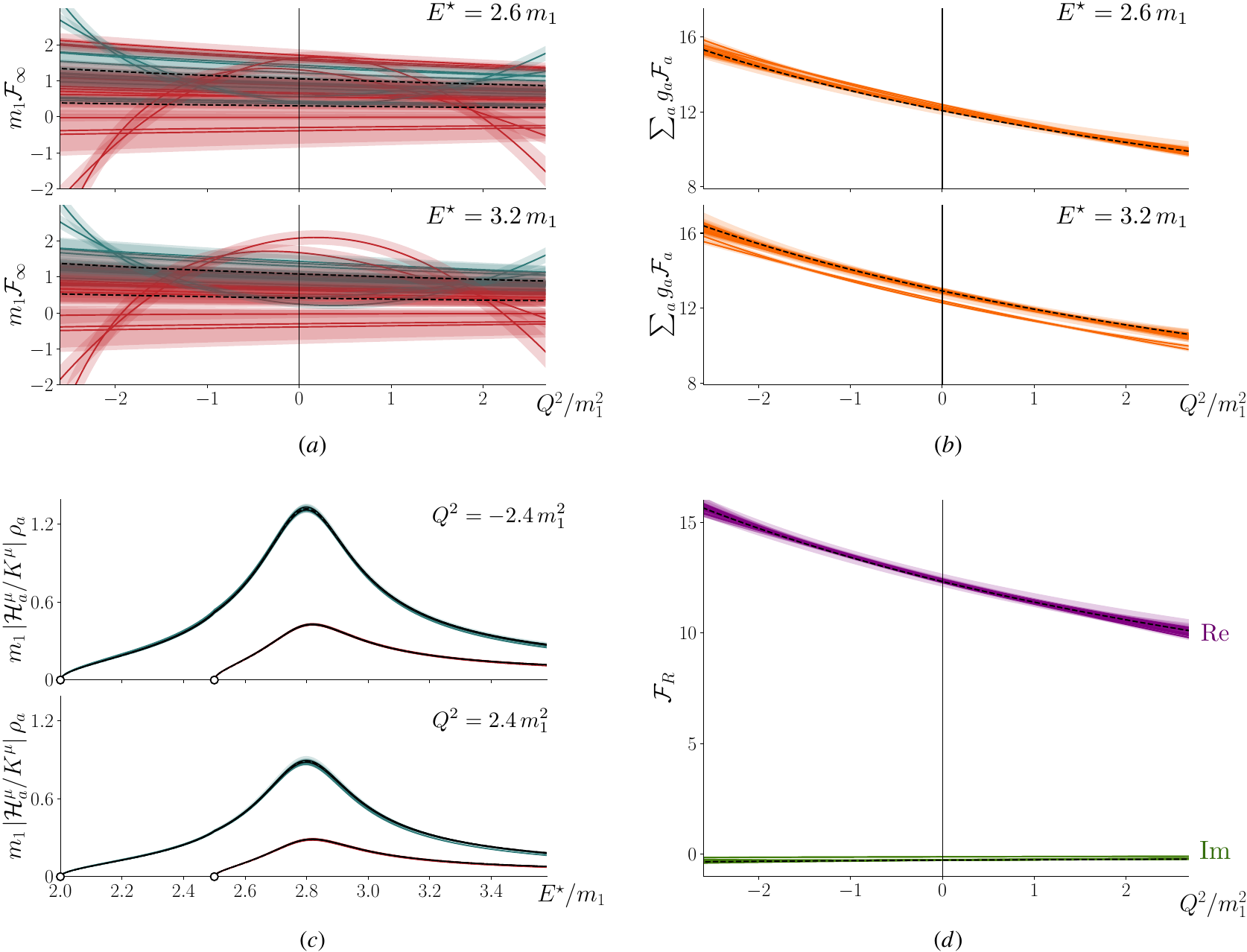}
  \caption{As Fig.~\ref{fig:Cusp_fits} but for the ``\emph{Flatt\'e-like}'' amplitude where there are 13 successful descriptions of the $\Fc_L$ synthetic data. }
  \label{fig:flatte_fits}
  \end{center}
\end{figure*}
%%%%%%%%%%%%%%%%%%%%%%%%%%%%%%%%%%%%%%%%%%%%%%%%%%%%%%%%%%%%%%%%%%

We retain all parameterizations found capable of describing the finite-volume matrix element data with a $\chi^2/N_\mathrm{dof}$ below a nominal~\footnote{Given our relatively simple approach to placing uncorrelated errors on our synthetic data, one should not assign too much meaning to the particular value of this cutoff.} cutoff of 2.5. The results are shown in Fig.~\ref{fig:Cusp_fits} for the ``\emph{cusp-like}'' amplitude and in Fig.~\ref{fig:flatte_fits} for the ``\emph{Flatt\'e-like}'' amplitude.

Fig.~\ref{fig:Cusp_fits} for the ``\emph{cusp-like}'' amplitude shows 11 successful descriptions which demonstrate that we can reliably reconstruct the transition process using just the limited set of matrix-element values on a single volume. This was perhaps to be expected given the relatively mild $s$-dependence and simple monotonic fall off in $Q^2$ of $\widetilde{\Fc}_a(Q^2,s)$. 

Fig.~\ref{fig:flatte_fits} for the ``\emph{Flatt\'e-like}'' amplitude, shows 13 successful descriptions, and is superficially similar to the ``\emph{cusp-like}'' case in panels (c) and (d), which show an accurate reconstruction of the transition amplitude and the form-factor at the resonance pole. On the other hand, panel (a) has the individual channel form-factors showing a high degree of scatter over parameterizations, to the extent that we cannot even make qualitative statements about their behavior. In fact we anticipated this as a feature of the near-rank-one nature of our ``\emph{Flatt\'e-like}'' amplitude, where only the combination $g_1 \Fc_1 + g_2 \Fc_2$ is well defined. This quantity is plotted in panel (c) and we see that it has a drastically reduced scatter over parameterizations compared with panel (a).

This exercise shows that, apart from a rather unique quirk of the ``\emph{Flatt\'e-like}'' amplitude, the set of finite-volume matrix elements we are considering is sufficient to reconstruct the underlying transition amplitude, making only mild assumptions about the behavior of the form-factors. However, this is a deliberately idealized situation in that we have assumed the scattering matrix $\Mc(s)$ to be known exactly, while in practical lattice QCD calculations this is not the case, as $\Mc(s)$ has to be determined by describing finite-volume energy spectra extracted from lattice QCD computed two-point correlation functions. We will now extend our synthetic data study to more closely resemble this.

\pagebreak
%%%%%%%%%%%%%%%%%%%%%%%%%%%%%%%%%
\subsection{Practical situation }\label{sec:varyM_varyF}
%%%%%%%%%%%%%%%%%%%%%%%%%%%%%%%%%

In this case we restrict our attention to the ``\emph{cusp-like}'' model, noting that the ``\emph{Flatt\'e-like}'' model gives similar results modulo the peculiarities arising from its near-rank-one nature. 

In order to mimic a realistic lattice QCD spectrum, we take the finite-volume spectrum coming from solution of the quantization condition on three volumes (${m_1  L = 3,4,5}$) in five frames. The exact spectra for our ``\emph{cusp-like}'' amplitude is shown as the orange curves in Figure~\ref{fig:spectra}(a), and for each discrete energy level on the three volumes we draw from a gaussian probability distribution such that the error on the energy is at the 5\% level with a commensurate random fluctuation of the mean. The resulting uncorrelated data are shown in Figure~\ref{fig:spectra}(a) as the points with errorbars.

This procedure provides us with 63 energy levels, and the selected set of volumes and frames is rather similar to the explicit lattice QCD calculations of coupled-channel scattering presented in Refs.~\cite{Briceno:2017qmb,Dudek:2016cru}. We have a comparable number of energy levels to those calculations, while our synthetic data errors are actually somewhat larger than those found for most levels therein.

%%%%%%%%%%%%%%%%%%%%%%%%%%%%%%%%%%%%%%%%%%%%%%%%%%%%%%%%%%%%%%%%%%
\begin{figure}[b!]
\includegraphics[width=.4\textwidth]{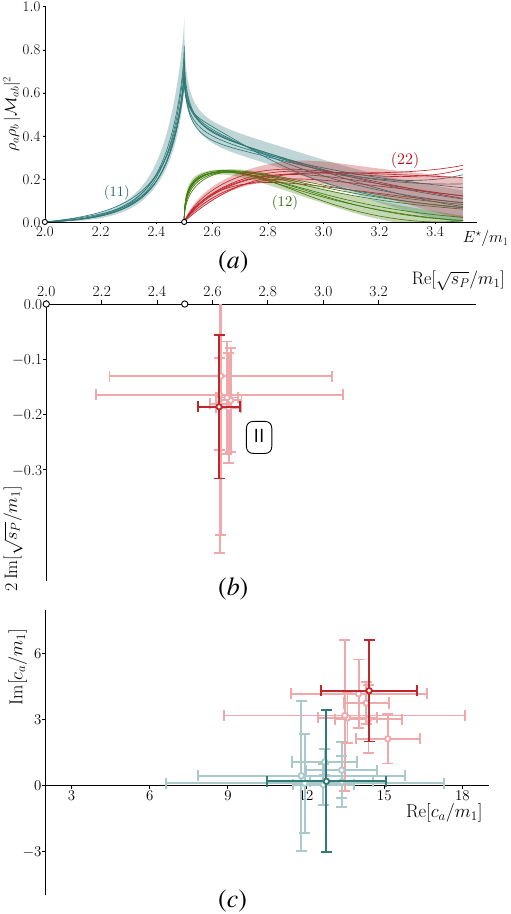}
  \caption{ Variation in the ``\emph{cusp-like}'' $\Mc$ when describing the finite-volume spectra by a range of parameterizations as described in the text. (a) Colored bands ($1\sigma$ variation) show the amplitude obtained using the correct $\Mc$ parameterization (the one used to generate the spectra), while curves show the central values of six other parameterizations. (b) Second sheet pole location. (c) Pole couplings to $\varphi_1 \varphi_1$(blue) and $\varphi_2 \varphi_2$(red). }
  \label{fig:M2_final_fits_L5}
\end{figure}
%%%%%%%%%%%%%%%%%%%%%%%%%%%%%%%%%%%%%%%%%%%%%%%%%%%%%%%%%%%%%%%%%%
 
With these energy levels in hand, we proceed assuming that we do not know the form of the underlying scattering matrix $\Mc(s)$, beyond that it satisfies coupled-channel unitarity. We propose a range of parameterizations, and by minimizing a $\chi^2$ for each one in an attempt to describe the spectra, obtain a set of plausible amplitudes. Seven such amplitudes are found describing the spectra with $\chi^2/N_\mathrm{dof} < 1.5$. They are all $K$-matrix forms -- several use Eq.~\eqref{eq:K} with some elements of the $\gamma$ matrix set to zero, others generalize Eq.~\eqref{eq:K} to use a polynomial in $s$ in place of $\gamma$, one uses two poles in $s$, and two make use of a form where the elements of $\Kc$ are expressed as a ratio of low-order polynomials.
 
Figure~\ref{fig:M2_final_fits_L5} shows the results of these amplitude applied to description of the synthetic finite-volume spectra. The bands ($1\sigma$ variation) show the amplitude obtained using the correct $\Mc$ parameterization (the one used to generate the spectra) and hence this reflects the best possible description of the scattering system given the noise on the finite-volume energy levels. The curves show the other parameterizations, where we observe that in the region where there is constraint from energy levels, the descriptions broadly agree, differing only at a level comparable to the statistical fluctuations on the correct amplitude. Also shown are the pole singularity location and couplings for each parameterization, which we observe to also be in quite reasonable agreement. These observations regarding the description of finite-volume spectra using a range of coupled-channel parameterizations are quite similar to those made in Ref.~\cite{Guo:2012hv}, and subsequently observed in several explicit lattice QCD calculations~\cite{Dudek:2014qha, Wilson:2014cna, Briceno:2017qmb, Dudek:2016cru, Woss:2019hse, Moir:2016srx, Woss:2020ayi}.

%%%%%%%%%%%%%%%%%%%%%%%%%%%%%%%%%%%%%%%%%%%%%

%%%%%%%%%%%%%%%%%%%%%%%%%%%%%%%%%%%%%%%%%%%%%%%%%%%%%%%%%%%%%%%%%%
\begin{figure}
\includegraphics[width=.48\textwidth]{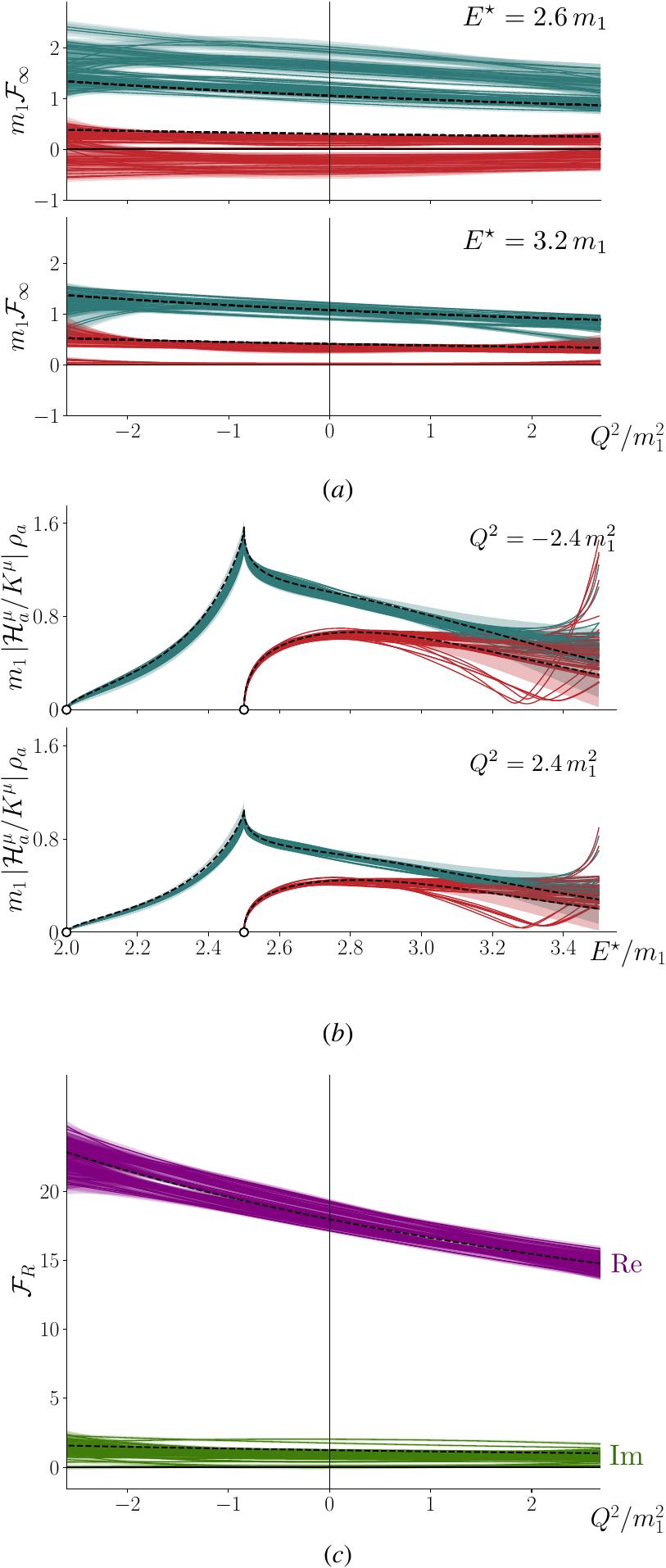}
  \caption{(a) Form-factors, (b) transition amplitudes, and (c) transition form-factor of the resonance, shown for 57 descriptions of the synthetic finite-volume spectra and finite-volume matrix element data, as described in the text.}
  \label{fig:Cusp_fits_M2var}
\end{figure}
%%%%%%%%%%%%%%%%%%%%%%%%%%%%%%%%%%%%%%%%%%%%%%%%%%%%%%%%%%%%%%%%%%

\vspace{3mm}
With a set of plausible $\Mc(s)$ forms, we can now repeat the analysis of the previous subsection, but this time propagating the parameterization variation of $\Mc(s)$ into the matrix element analysis. Practically, we generate the ${\mu_0^\star}'$ and $\mathbf{w}_0$ in Eq.~\eqref{eq:cdotMap} for each of the plausible $\Mc(s)$ models, and for each one consider in addition the variations of $\Fc_a(Q^2, s)$ parameterization detailed previously. Doing so in a description of the synthetic $\Fc_L$ data we find  57 combinations which have a $\chi^2/N_\mathrm{dof}$ below a nominal cutoff of 2.5. The resulting amplitudes are shown in Fig.~\ref{fig:Cusp_fits_M2var} where we observe, in comparison to Fig.~\ref{fig:Cusp_fits}, a somewhat larger spread in $\Fc_{1,2}$ curves, but still relatively little variation in the transition amplitudes, $\Hc_{1,2}$, in the energy region where there is data providing constraint. We suggest that this is due to the fact that \emph{on the real energy axis}, the polynomial behavior in $s$ of the form-factors can compensate for erroneous slow $s$-dependence in $\Mc$ caused by use of an imperfect parameterization. Given this hypothesis we might expect the resonance form-factor, evaluated at the pole in the complex energy plane (Eq.~\ref{eq:resff}), to show a larger degree of parameterization variation than $\Hc_{1,2}$, since the $s$-dependence ``compensation'' described above is only constrained on the real energy axis. Indeed this is what we observe in Fig.~\ref{fig:Cusp_fits_M2var}(c), but the degree of fluctuation is modest, and it is clear that the $Q^2$ behavior of the dominant real part is reproduced, as is the large hierarchy with respect to the imaginary part.

Similar to the spectral analysis leading to $\Mc(s)$, one observes that the transition amplitudes are more poorly constrained at higher energies. This can be easily understood by revisiting the synthetic data for the spectrum shown in Fig.~\ref{fig:spectra}(a), where one sees that for the $L=5/m_1$ volume, there is only a single energy level above $E^\star =3.2\, m_1$ providing constraint.

The precise degree of scatter observed is a function of the $\chi^2$ cutoff imposed on the descriptions of the \mbox{finite-volume} spectra and the finite-volume matrix element data. As this is sensitive to how one generates the noise on the synthetic data, and since we chose a rather simple approach, we selected a rather loose cutoff. More careful consideration of the statistical behavior will be justified when real lattice QCD is in hand. But given this slight caveat, this second analysis does expose the importance of considering the systematic uncertainty arising from the description of the scattering matrix when performing calculations of transition matrix elements.

%%%%%%%%%%%%%%%%%%%%%%%%%%%%%%%%%%%%
% Section V - summary
%%%%%%%%%%%%%%%%%%%%%%%%%%%%%%%%%%%%
\section{Summary}\label{sec:V}

In this work we have presented a first investigation of the implementation of the formalism derived in References~\cite{Briceno:2014uqa, Briceno:2015csa} for studying $1 \xrightarrow{\mathcal{J}} 2$ transition processes where the final state can be one of several open channels. We have rewritten the generalized Lellouch-L\"uscher matrix using an eigenvalue decomposition, which provides a relatively simple conceptual picture, where the \mbox{finite-volume} hadron-hadron states are normalized by a factor featuring the slope of the eigenvalues, while the channel admixture is provided by the eigenvectors. 

We have explained how a Lorentz decomposition for these transition amplitudes can be performed, parametrizing the dynamics of such processes in terms of Lorentz scalar functions that can be understood as energy-dependent form factors. We have performed this decomposition explicitly for the simplest non-trivial case, where the current is a conserved Lorentz vector and the initial and final states are scalars. 

We built a pair of toy-models each featuring a single resonance coupled to two meson-meson scattering channels, and demonstrated that even with a realistically limited number of matrix-element values with reasonable uncertainties, one can extract the transition amplitudes, and also place significant  constraints on the resonance transition form-factors though analytic continuation of the amplitudes into the complex energy plane.

Within this toy-model analysis we observed a systematic uncertainty in the transition process which arises from parameterization variation when describing the scattering amplitude, constrained by the finite-volume spectrum. For kinematical regions where there are significant numbers of synthetic spectrum points and matrix elements, the systematic error in the amplitudes are comparable to the statistical, while in kinematical regions where there are fewer constraints, the systematic errors due to the parametrization choice is likely to dominate the error budget.  

The toy-model examples considered here are expected to be most immediately relevant for transition amplitude studies involving the $a_0$ resonance. In particular, two phenomenologically interesting processes that could be studied using these techniques are $\gamma \omega \to (\eta \pi, K\overline{K})$ and $\gamma \phi \to (\eta \pi, K\overline{K})$ where the coupled system in the final state features the $a_0$ resonance in $S$-wave. In a previous lattice QCD determination of the scattering amplitude, Ref.~\cite{Dudek:2016cru}, a result rather similar to the ``\emph{cusp-like}'' model was found. By studying the $\omega \to a_0$ and $\phi \to a_0$ resonance transition form factors, identifying $\omega$ and $\phi$ as being predominantly of $\ell\overline{\ell}$ and $s\overline{s}$ construction respectively, one can begin to explore the internal quark flavor structure of the $a_0$.

Extensions of these ideas for baryonic systems like $\gamma N \to \pi N, \eta N$ is feasible~\cite{Briceno:2015csa}. The Lorentz decomposition and cubic-subduction is more complex than the scenario considered in this paper, but the extension is a straightforward application of known methods. The main practical challenge for phenomenologically interesting reactions is that three-hadron thresholds open in the energy region of interest. For example, for physical values of the quark masses, transitions coupling $\pi N$ and $\eta N$ in the final state will also couple to $ \pi\pi N$. Three-body states present a new class of challenges, but in recent years there has been tremendous formal process to understand the spectrum of three-particle systems~\cite{Hansen:2014eka, Hansen:2015zga,Briceno:2018mlh, Briceno:2017tce, Jackura:2020bsk} which has resulted in the first determination of a three-body scattering amplitude from lattice QCD~\cite{Hansen:2020otl}. Most recently the first step towards generalizing the Lellouch-L\"uscher matrix for kinematics where three-body systems can go on-shell has been presented~\cite{Hansen:2021ofl}. As a result, it is not unreasonable to expect these ideas to be extended in the upcoming years to accommodate mixing between two- and three-particle states. 

Finally, we comment that similar analysis techniques to those proposed in this paper will be necessary in the implementation of the already existing formalism for $2 \xrightarrow{\mathcal{J}} 2$ reactions~\cite{Baroni:2018iau, Briceno:2015tza, Briceno:2020xxs}. The finite-volume formalism, which has already gone through rigorous formal testing~\cite{Briceno:2019nns, Briceno:2020xxs}, when combined with an understanding of the analytic structure of the subsequent amplitudes~\cite{Briceno:2020vgp} will provide access to the elastic form-factors of narrow hadron resonances. Such quantities provide a set of novel observables, not accessible in experiment, which can inform our understanding of the internal structure of unstable excited hadrons. 

%%%%%%%%%%%%%%%%%%%%%%%%%%%%%%%%%%%%%%%%%%%%%%%%%%%%%%%%%%%%%%%%%%%%%%%%%%%%%%%%%%%%%%%%%%%%%%%%%%%%%%%%%%%%%%%%%%%%%%%%%%%%%%%%

\vspace*{5mm}
\section{Acknowledgements}
The authors acknowledge Ben Slimmer for his assistance in the early stages of this work. The authors thank Andrew Jackura for useful comments on an early version of the manuscript. RAB and JJD are supported in part by U.S. Department of Energy Contract No. DE-AC05-06OR23177, 
under which Jefferson Science Associates, LLC, manages and operates Jefferson Lab. JJD acknowledges support from the U.S. Department of Energy Contract No. DE-SC0018416. RAB and LL acknowledge support from the U.S. Department of Energy Contract No. DE-SC0019229. 

\appendix
%%%%%%%%%%%%%%%%%%%%%%%%%%%%%%%%%%%%
% Appendix A - properties of R
%%%%%%%%%%%%%%%%%%%%%%%%%%%%%%%%%%%%
\section{Properties of $\widetilde{\mathcal{R}}_n$} \label{app:R}

The matrix whose eigenvalues and eigenvectors we need in order to construct $\widetilde{\mathcal{R}}_n$ is ${F(E^\star,\mathbf{P}; L) + \Mc^{-1}(E^\star)}$. This features $F(E^\star,\mathbf{P}; L)$ which houses `kinematic' \mbox{finite-volume} functions and which is diagonal in \mbox{channel-space}, but in general has entries connecting different partial-waves. Its definition, and the technology to \emph{subduce} it into irreducible representations of cubic symmetry can be found in the appendix of Ref.~\cite{Briceno:2017max}. In short, the elements of the matrix $F$ subduced into irrep $\Lambda$ are of the form
\begin{equation*}
 F^\Lambda_{\ell n; \ell' n'} = i \rho \big( \delta_{\ell \ell'} \delta_{n n'} + i f^\Lambda_{\ell n; \ell' n'} \big)
\end{equation*}
where the embedding label $n$ is required in cases where $\ell$ subduces more than once into $\Lambda$.

%%%%%%%%%%%%%%%%%%%%%%%%%%%%%%%%%%%%%%%%%%%%%%%%%%%%%%%%%%%%%%%%%%%%%%%%%
\subsection{Properties of $F + \Mc^{-1}$ for a single partial-wave}

We can illustrate some properties of the matrix $F + \Mc^{-1}$ using the example of two coupled-channels in a single partial wave of angular momentum $\ell$. The phase-space $\rho$ and the finite-volume functions $f$ are real above kinematic threshold for each channel, and imaginary below. Hence for energies lying above both kinematic thresholds, assuming time-reversal symmetry such that $\Mc$ is symmetric,
\begin{widetext}
\begin{align*}
F + \Mc^{-1} &= 
\begin{bmatrix}
i \rho_1 \big(1 + i f_1 \big) & 0 \\
0 & i \rho_2 \big(1 + i f_2 \big)
\end{bmatrix}
+ 
 \begin{bmatrix}
-i \rho_1 + \mathrm{Re} \big(\Mc^{-1}\big)_{11} & \mathrm{Re} \big(\Mc^{-1}\big)_{12} \\
\mathrm{Re} \big(\Mc^{-1}\big)_{12} & -i \rho_2 + \mathrm{Re} \big(\Mc^{-1}\big)_{22}
\end{bmatrix} \nonumber \\
&= \begin{bmatrix}
- \rho_1 f_1 + \mathrm{Re} \big(\Mc^{-1}\big)_{11} 
& \mathrm{Re} \big(\Mc^{-1}\big)_{12} \\
\mathrm{Re} \big(\Mc^{-1}\big)_{12} 
& - \rho_1 f_1 + \mathrm{Re} \big(\Mc^{-1}\big)_{22}
\end{bmatrix} \, ,
\end{align*}
where use has been made of the unitarity condition, Eq.~\eqref{eq:unitarity}. Clearly $F + \Mc^{-1}$ is \emph{real} and \emph{symmetric}, ensuring that its eigenvectors are \emph{orthogonal}, $\mathbf{w}_i^\intercal\!\cdot\! \mathbf{w}_j = \delta_{ij}$. 

\vspace{3mm}
In practice we may have to evaluate $\widetilde{\mathcal{R}}_n$ at a finite-volume energy which lies above threshold for some channels, but below the threshold for others. In our two-channel illustration we can consider the energy region above the threshold for channel 1, but below the threshold for channel 2. In that case,
\begin{align*}
F + \Mc^{-1} &= 
\begin{bmatrix}
i \rho_1 \big(1 + i f_1 \big) & 0 \\
0 & i (i \hat{\rho}_2) \big(1 + i (i \hat{f}_2) \big)
\end{bmatrix} 
+
\begin{bmatrix}
-i \rho_1 + \mathrm{Re} \big(\Mc^{-1}\big)_{11} & \mathrm{Re} \big(\Mc^{-1}\big)_{12} \\
\mathrm{Re} \big(\Mc^{-1}\big)_{12} & \mathrm{Re} \big(\Mc^{-1}\big)_{22}
\end{bmatrix} \nonumber \\
&= \begin{bmatrix}
- \rho_1 f_1 + \mathrm{Re} \big(\Mc^{-1}\big)_{11} 
& \mathrm{Re} \big(\Mc^{-1}\big)_{12} \\
\mathrm{Re} \big(\Mc^{-1}\big)_{12} 
& \hat{\rho}_2 \big( \hat{f}_2 - 1 \big) + \mathrm{Re} \big(\Mc^{-1}\big)_{22}
\end{bmatrix} \, ,
\end{align*}
where $\hat{\rho}_2, \hat{f}_2$ are real functions. It is clear that $F + \Mc^{-1}$ is still \emph{real} and \emph{symmetric} and its eigenvectors remain \emph{orthogonal}. 
\end{widetext}

An interesting case is when we remain above threshold for channel 1, but are \emph{far below} threshold for channel 2, as here we would expect the physics of scattering in \mbox{channel 2} to become irrelevant. The property of the finite-volume functions required here is that far below the threshold for channel $a$, $f_a \to i$, and hence $F_a \to 0$. It follows that
\begin{equation*}
F + \Mc^{-1} \to 
 \begin{bmatrix}
- \rho_1 f_1 + \mathrm{Re} \big(\Mc^{-1}\big)_{11} 
& \mathrm{Re} \big(\Mc^{-1}\big)_{12} \\
\mathrm{Re} \big(\Mc^{-1}\big)_{12} 
&  \mathrm{Re} \big(\Mc^{-1}\big)_{22}
\end{bmatrix} \, ,
\end{equation*}
and if we compute the determinant of this matrix we obtain
\begin{equation*}
 \det \big(F + \Mc^{-1} \big) = \frac{\Mc_{11} }{\det \Mc } \left[ i \rho_1 \big( 1 + i f_1 \big) + \frac{1}{\Mc_{11} } \right],
\end{equation*}
where we recognize the object in square brackets as the corresponding quantization condition if only channel 1 existed and not channel 2. So, as expected, far below the threshold for channel 2, the finite-volume spectrum is controlled only by channel 1. It is straightforward to show that in this case the relevant eigenvector at the finite-volume energy is
\begin{equation*}
\mathbf{w}_0 \propto 
\begin{bmatrix}
 \big(\Mc^{-1}\big)_{22} \\
 -  \big(\Mc^{-1}\big)_{12} \\
\end{bmatrix} \, , 
\end{equation*}
but because the combination $\Mc^{-1} \mathbf{w}_0$ appears in the construction of $\widetilde{\mathcal{R}}_n$, Eq.~\eqref{eq:altR}, and
\begin{align*}
\Mc^{-1} \mathbf{w}_0 &\propto 
\begin{bmatrix}
 \big(\Mc^{-1}\big)_{11} &  \big(\Mc^{-1}\big)_{12}\\
  \big(\Mc^{-1}\big)_{12} & \big(\Mc^{-1}\big)_{22} \\
\end{bmatrix}
\begin{bmatrix}
 \big(\Mc^{-1}\big)_{22} \\
 -  \big(\Mc^{-1}\big)_{12} \\
\end{bmatrix} \\
&= \begin{bmatrix}
\det \Mc^{-1} \\
0
\end{bmatrix} ,\,
\end{align*}
we see that channel 2, as anticipated, decouples completely from the problem.

%%%%%%%%%%%%%%%%%%%%%%%%%%%%%%%%%%%%%%%%%%%%%%%%%%%%%%%%%%%%%%%%%%%%%%%%%
\subsection{Properties of $F + \Mc^{-1}$ for multiple partial-waves}

While $F + \Mc^{-1}$ is always a symmetric matrix, it is not guaranteed in general that all elements are real. An illustration is provided by the case of a single scattering channel in which the scattering particles have differing masses and can scatter in two partial-waves. Considering the $\ell=0\, (S)$ and $\ell=1\, (P)$ partial waves in the $[110]\, A_1$ irrep, the relevant $f$-functions \emph{below} the kinematic threshold are shown in Fig.~26 of Ref.~\cite{Briceno:2017max}. We observe that $f_{SS}$ and $f_{PP}$ are imaginary and positive, while $f_{SP}$ is real and positive. Since the phase-space below threshold is imaginary and positive, $\rho = i \hat{\rho}$, and the diagonal $\Mc^{-1}$ is real below threshold, we have
\begin{equation*}
	F + \Mc^{-1} = 
	\begin{bmatrix}
		- \hat{\rho} \big( 1 - \hat{f}_{SS} \big) + \tfrac{1}{\Mc_{S}} & -i \hat{\rho} f_{SP} \\
		-i \hat{\rho} f_{SP} & - \hat{\rho} \big( 1 - \hat{f}_{PP} \big) + \tfrac{1}{\Mc_{P}}
	\end{bmatrix} \, ,
\end{equation*}
which is symmetric but not real. The symmetry is sufficient to ensure that the eigenvectors are orthogonal, $\mathbf{w}_i^\intercal \!\cdot\! \mathbf{w}_j = \delta_{ij}$, but note that we must unit normalize with the \emph{transpose} and not the hermitian conjugate. In fact from the form of the matrix having real diagonal elements and imaginary off-diagonal elements, it is clear that the normalized eigenvector having zero eigenvalue will take the form
\begin{equation*}
\mathbf{w}_0 = \frac{1}{\sqrt{b^2 - a^2}} \begin{bmatrix} b \\ i\,  a \end{bmatrix},
\end{equation*}
where we see that the $P$-wave component is imaginary. We might worry that this will cause a problem in Eq.~\eqref{eq:cdotMap} as this should be a real-valued matrix-element. In fact there is no problem because of the required factor $\tfrac{1}{q^\star}$ for a $P$-wave, which below kinematic threshold provides a compensating factor of $i$.

\vspace{3mm}
Within this illustrative example, we can also consider how the presence of a \emph{deeply-bound bound-state} would manifest in a finite-volume. Far below the kinematic threshold, as can be seen in Fig. 26 of Ref.~\cite{Briceno:2017max}, ${\hat{f}_{SS} \to 1}$, ${\hat{f}_{PP} \to 1}$, ${f_{SP} \to 0}$, such that ${F+\Mc^{-1} \to \Mc^{-1}}$ and the dependence on the finite-volume disappears. Since $\Mc$ is diagonal in partial-waves, the eigenvalues are trivially $\tfrac{1}{\Mc_{S}}, \tfrac{1}{\Mc_P}$ with eigenvectors $\begin{bsmallmatrix}1 \\ 0\end{bsmallmatrix}$, $\begin{bsmallmatrix}0 \\ 1\end{bsmallmatrix}$ respectively.\\[-0.9ex]

If we place a deeply-bound pole in $S$-wave by writing $\Mc_S(E^\star) = \frac{g^2}{m_\mathrm{bs}^2 - E^{\star 2}}$, then the finite-volume energy will be at $E^\star_n = m_\mathrm{bs}$, and
\begin{align*}
 \widetilde{\mathcal{R}}_n &= \left( - \tfrac{2m_\mathrm{bs}}{{\mu_0^\star}'} \right) \Mc^{-1}  \mathbf{w}_0 \, \mathbf{w}_0^\intercal \, \Mc^{-1} \\
 &= \frac{ \big( m_\mathrm{bs}^2 \!-\! E^{\star 2} \big)^2}{g^2} \begin{bmatrix} 1 & 0 \\ 0 & 0  \end{bmatrix}.
\end{align*}
Then using Eq.~\eqref{eq:LL_formula} we have
\begin{equation*}
 \big\langle \mathcal{J} \big\rangle = \frac{1}{L^3 \sqrt{2 E_i} \sqrt{2 E_f}} \, \frac{ m_\mathrm{bs}^2 \!-\! E^{\star 2} }{g} \,  \mathcal{H}_S \, ,
\end{equation*}
and it is natural to write $\mathcal{H}_S = \frac{ g \,h }{m_\mathrm{bs}^2 - E^{\star 2}}$ where $h$ is interpreted as the coupling for $(\gamma i \to \mathrm{bs})$ so
\begin{equation*}
\big\langle \mathcal{J} \big\rangle = \frac{1}{L^3 \sqrt{2 E_i}\sqrt{2E_f}} \, h \, ,
\end{equation*}
which is what we would expect for the transition, induced by the current, from stable single particle $i$, to stable single particle $(\mathrm{bs})$, where $E_f = \sqrt{m_\mathrm{bs}^2 + \mathbf{P}_{\!f}^2}$.

%%%%%%%%%%%%%%%%%%%%%%%%%%%%%%%%%%%%%%%%%%%%%%%%%%%%%%%%%%%%%%%%%%%%%%%%%
\subsection{Zero-crossing eigenvalues of $F + \Mc^{-1}$}

\begin{figure}[h]
  \begin{center}
\includegraphics[width=.48\textwidth]{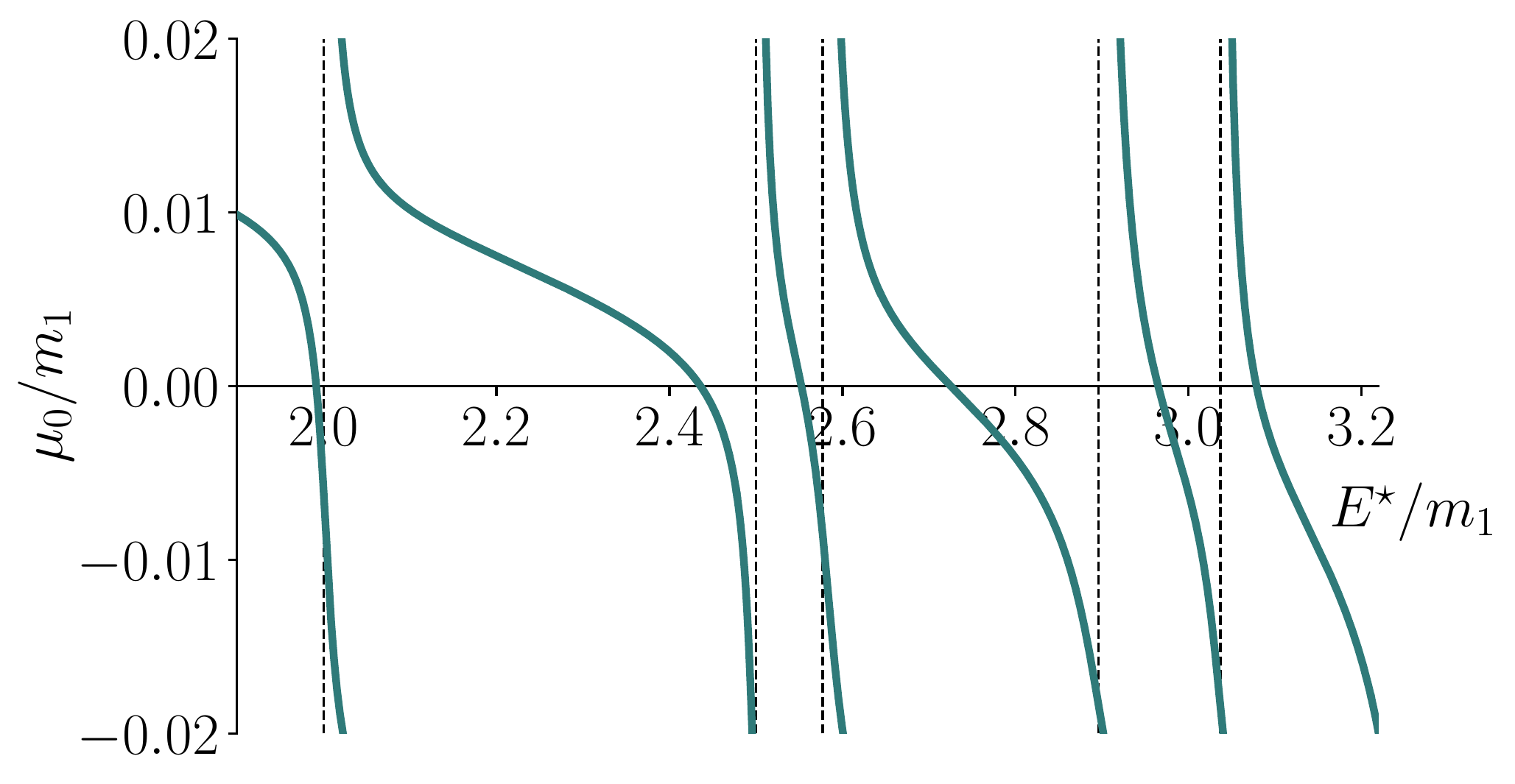}
  \caption{The zero-crossing eigenvalue of $F + \Mc^{-1}$, evaluated for the ``\emph{cusp-like}'' amplitude, in the irrep $[002] A_1$ with ${L = 6/m_1}$. The vertical dashed lines indicate the non-interacting energies for this system. The other eigenvalue of this two channel system takes values of much larger magnitude than the scale of this plot.}
  \label{fig:mu_plot}
  \end{center}
\end{figure}

The finite-volume normalization factor appearing in Eq.~\eqref{eq:cdotMap}, $\sqrt{-\tfrac{2E^\star_n}{{\mu_0^\star}'}}$ requires that the slope of the zero crossing eigenvalue of $F + \Mc^{-1}$ at $E^\star_n$ must be negative. Indeed as shown in Fig.~\ref{fig:mu_plot}, which illustrates the case of the ``\emph{cusp-like}'' amplitude discussed in this paper, the zero crossing eigenvalue of $F + \Mc^{-1}$ falls-off monotonically on the intervals between non-interacting energies and crosses zero with a negative slope. This proves to be the case for every amplitude we have considered, and is presumably a general property.

%%%%%%%%%%%%%%%%%%%%%%%%%%%%%%%%%%%%%%%%%%%%%%%%%%%%%%%%%%%%%%%%%%
%%%%%%%%%%%%%%%%%%%%%%%%%%%%%%%%%%%%%%%%%%%%%%%%%%%%%%%%%%%%%%%%%%

%%%%%%%%%%%%%%%%%%%%%%%%%%%%%%%%%%%%%%%%%%%%%%%%%%%%%%%%%%%%%%%%%%%%%%%%%
\pagebreak
\subsection{$\widetilde{\mathcal{R}}$--matrix away from finite-volume energies}

An approach which has been followed in past elastic calculations~\cite{Briceno:2016kkp, Briceno:2015dca, Alexandrou:2018jbt} is generalize the factor $\mathcal{R}_n$ such that it is a continuous function of energy. This can be seen in Eqs.~(19) and (20) of Ref.~\cite{Briceno:2015dca} where the variable $r(E)$ takes value $1$ at the energies $E_n$ satisfying the quantization condition, but varies from $1$ away from these energies. The motivation for this choice was to be able to evaluate the finite-volume normalization factor at the actual computed lattice QCD energies. In fact, this is not a unique procedure, and can lead to an uncontrolled systematic error, as we will now illustrate for the more general case of coupled-channel transition amplitudes.

\vspace{3mm}
Suppose we take the defining equation for $\widetilde{\mathcal{R}}_n$,
\begin{equation*}
\widetilde{\mathcal{R}}_n(\mathbf{P}, L) \equiv 2E_n \cdot \lim_{E \to E_n} (E - E_n) \Big( F^{-1}(E^\star,\mathbf{P}; L) + \Mc(E^\star) \Big)^{-1} \, ,
\end{equation*}
and consider there to be a generalization,
\begin{equation*}
\widetilde{\mathcal{R}}(E, \mathbf{P}, L) \equiv 2E_n \cdot\, (E - E_n) \Big( F^{-1}(E^\star,\mathbf{P};L) + \Mc(E^\star) \Big)^{-1} \, ,
\end{equation*}
valid in an energy region around each $E_n$, and which is equal to $\widetilde{\mathcal{R}}_n$ when $E = E_n$. We might consider this to be a way to obtain the ``Lellouch-L\"uscher'' factor at lattice QCD energies, even when the scattering model in finite volume does not exactly match those energies ($E_\mathrm{lat.} \neq E_n$).

\begin{figure}[h]
  \begin{center}
  \includegraphics[width=.48\textwidth]{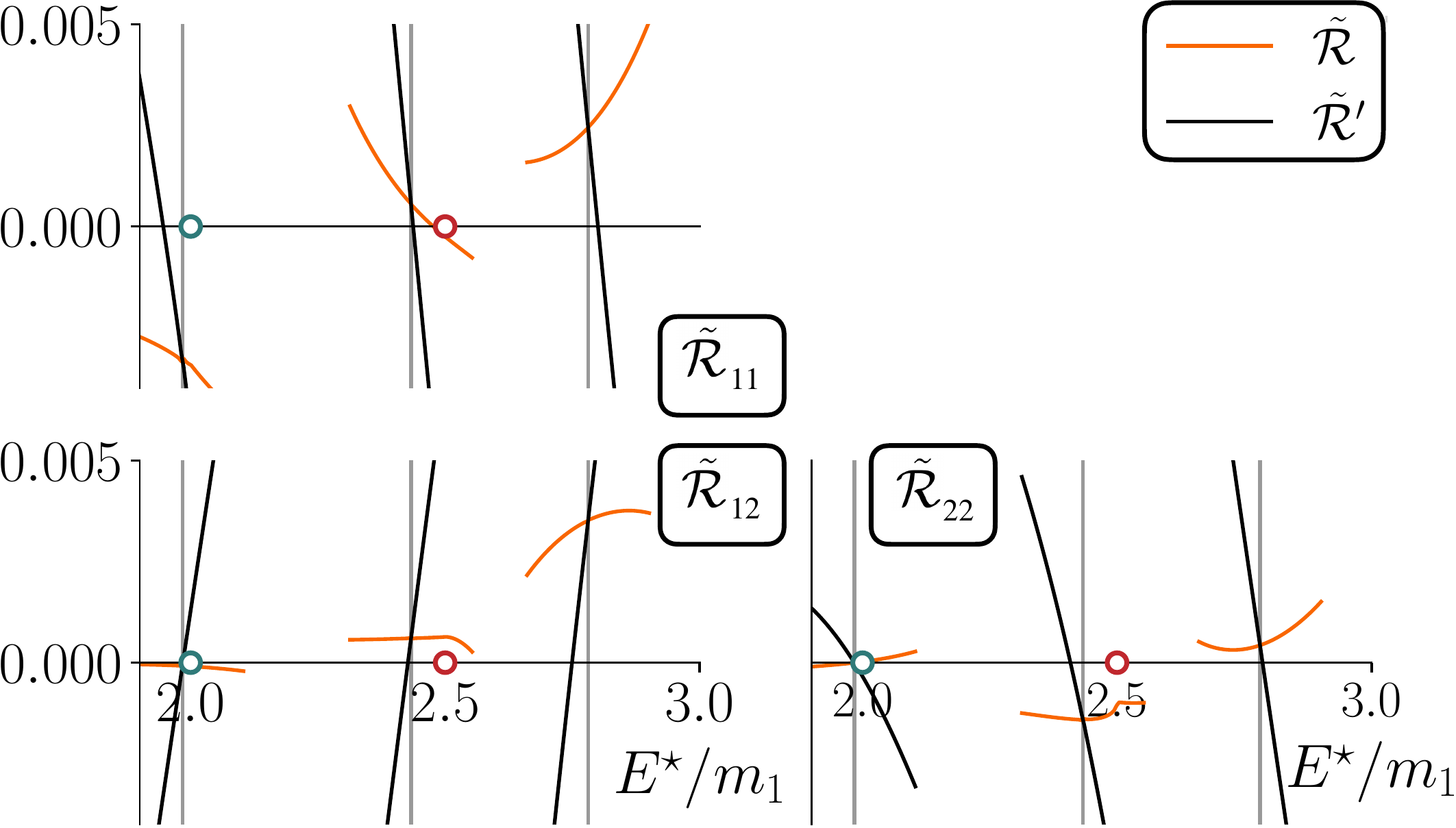}
  \caption{ $\widetilde{\mathcal{R}}$ and  $\widetilde{\mathcal{R}}'$ as defined in the text calculated with the ``\emph{cusp-like}'' amplitude in the $[000]A_1$ irrep in a volume $L = 5/m_1$. The vertical lines indicate the locations of the finite-volume spectrum, $E^\star_n$.
  }
  \label{fig:R_Rprime}
  \end{center}
\end{figure}

One immediate issue with this is that there is not a \emph{unique} matrix function with this property -- for example, 
\begin{equation*}
\widetilde{\mathcal{R}}' \equiv - 2E_n \cdot (E - E_n) \, \Mc^{-1} \big( F + \Mc^{-1} \big)^{-1} \Mc^{-1} \, ,
\end{equation*}
is also equal to $\widetilde{\mathcal{R}}_n$ when $E = E_n$. This follows from the fact that at this energy,
\begin{equation*}
\widetilde{\mathcal{R}}_n = \frac{2E_n^\star}{ {\lambda_0^\star}'} \mathbf{v}_0 \mathbf{v}_0^\intercal
\end{equation*}
and because at this energy $\big[ F^{-1} + \Mc \big]\, \mathbf{v}_0 = 0$,
\begin{equation*}
F^{-1} \mathbf{v}_0 = - \Mc \mathbf{v}_0\,, \;\;\;
\mathbf{v}_0 = - F \Mc \mathbf{v}_0 \,, \;\;\;  \mathbf{v}_0^\intercal = \mathbf{v}_0^\intercal ( - \Mc F)  \,.
\end{equation*}
The last of these expressions makes it clear that there are an infinite number of such variations on $\widetilde{\mathcal{R}}$ possible, generated by left multiplying with arbitrary powers of $(- F \Mc)$ or right multiplying by powers of $(- \Mc F)$. Focussing on just $\widetilde{\mathcal{R}}'$ and $\widetilde{\mathcal{R}}'$ is sufficient to illustrate the point, that as shown in Figure~\ref{fig:R_Rprime}, these matrices can differ \emph{significantly} at energies away from $E=E_n$. In addition, the property that $\widetilde{\mathcal{R}}_n$ is rank-one does not hold for $\widetilde{\mathcal{R}}$ at generic energy values away from $E = E_n$, and as such the conceptually vital property of factorization which allows Eq.~\eqref{eq:cdotMap} to be extracted from Eq.~\eqref{eq:LL_formula} is removed.

%%%%%%%%%%%%%%%%%%%%%%%%%%%%%%%%%%%%%%%%%%%%%%%%%%%%%%%%%%%%%%%%%%
%%%%%%%%%%%%%%%%%%%%%%%%%%%%%%%%%%%%%%%%%%%%%%%%%%%%%%%%%%%%%%%%%%

This observation has an impact in practical lattice QCD calculations in the following way: in the approach we have proposed in this paper, the Lellouch-L\"uscher factor is only ever evaluated at finite-volume energies, $E^\star_n$, which correspond to solutions of the quantization condition for the parameterization of $\Mc$ being used to describe the lattice QCD spectrum data. In this case there is no ambiguity in the definition, as only $\widetilde{\mathcal{R}}_n$ ever appears. Of course these energies for the parameterized amplitude will typically not be exactly equal to the computed lattice QCD energies, rather as a set they form the best overall description of the spectrum under a $\chi^2$ minimization. An apparently appealing alternative approach is to evaluate $\tilde{\mathcal{R}}(E)$ \emph{at the lattice QCD energies}, but one sees immediately that in this case the problem of selecting a particular representation of $\widetilde{\mathcal{R}}$ arises, and the value of the finite-volume correction depends explicitly upon that choice. Furthermore, the ambiguity in the choice of representation of $\widetilde{\mathcal{R}}$ affects the propagation of uncertainty from the spectrum energies into the transition amplitude $\Hc$ such that the uncertainties of the observable quantity will depend (unreasonably) upon which representation of $\widetilde{\mathcal{R}}$ is chosen.

%%%%%%%%%%%%%%%%%%%%%%%%%%%%%%%%%%%%%%%%%%%%%%%%%%%%%%%%%
%%%%%%%%%%%%%%%%%%%%%%%%%%%%%%%%%%%%%%%%%%%%%%%%%%%%%%%%%
%%%%%%%%%%%%%%%%%%%%%%%%%%%%%%%%%%%%%%%%%%%%%%%%%%%%%%%%%
\vspace*{5mm}
\section{Flatt\'e amplitude in a finite-volume} \label{app:Flatte}

A scattering matrix of Flatt\'e type in $N$-channels can be written,
\begin{equation*}
\Mc^\mathrm{Fl.}(E^\star) = \mathbf{g} \frac{1}{D(E^\star)} \mathbf{g}^\intercal \, ,
\end{equation*}
where the real-valued couplings to each channel appear in a vector ${\mathbf{g}^\intercal = \big( g_1, g_2, \ldots, g_N \big)}$ and where ${D(E^\star) = m^2 - E^{\star 2} - i \sum_a g_a^2 \,\rho_a(E^\star)}$. Trivially we observe this matrix has one non-zero eigenvalue ${\tfrac{\mathbf{g}^\intercal \!\cdot \mathbf{g}}{D(E^\star)}}$ with eigenvector $\mathbf{g}$, and $N-1$ zero eigenvalues with eigenvectors orthogonal to $\mathbf{g}$.

In a finite volume, the quantization condition ${\det \big[ F + \mathcal{M}^{-1} \big] = 0}$ can equivalently be written $\det \big[ \mathcal{M} F + 1 \big] = 0$, and for the Flatt\'e amplitude,
\begin{equation*}
 \Mc^\mathrm{Fl.} F +1 = \frac{1}{D(E^\star)} \Big( \mathbf{g} \,\mathbf{g}^\intercal F(E^\star,\mathbf{P};L) + D(E^\star) \Big)  \, ,
\end{equation*}
and hence
\begin{equation*}
 \mathbf{g}^\intercal \big( \Mc^\mathrm{Fl.} F +1 \big) \mathbf{g}  = \frac{\mathbf{g}^\intercal \!\cdot\! \mathbf{g} }{D(E^\star)} \Big( \mathbf{g}^\intercal F(E^\star,\mathbf{P};L) \mathbf{g} + D(E^\star) \Big)  \, .
\end{equation*}
It follows that if we perform an orthogonal transformation on the matrix $\Mc^\mathrm{Fl.} F +1$ using a basis of vectors given by $\mathbf{g}$ and $N-1$ vectors orthogonal to $\mathbf{g}$ (and each other), we will obtain
a matrix which is the identity apart from one diagonal element which takes the value $\frac{\mathbf{g}^\intercal \!\cdot \mathbf{g} }{D(E^\star)} \Big( \mathbf{g}^\intercal F(E^\star,\mathbf{P};L) \mathbf{g} + D(E^\star) \Big)$. Upon taking the determinant it is clear that the finite-volume spectrum is given by solutions of 
\begin{equation*}
\sum\nolimits_a g_a^2 \,F_{aa}(E^\star,\mathbf{P};L) = - D(E^\star) \, .
\end{equation*}

For the corresponding zero eigenvalue of $F + \mathcal{M}^{-1}$,  $\big(\mathcal{M} F + 1 \big) \mathbf{w}_0 = 0$ so 
\begin{equation*}
 \big( \mathbf{g} \,\mathbf{g}^\intercal F + D \big) \mathbf{w}_0 = 0 \, ,
\end{equation*}
and given the condition for a zero eigenvalue, ${D(E^\star) = -\mathbf{g}^\intercal F(E^\star,\mathbf{P};L) \mathbf{g}}$, we have
\begin{equation*}
 \big( \mathbf{g} \,\mathbf{g}^\intercal F -\mathbf{g}^\intercal F \mathbf{g}  \big) \mathbf{w}_0 = 0 \, ,
\end{equation*}
which by inspection is solved by $\mathbf{w}_0 = \mathbf{g}$, manifestly independent of the particular energy level under consideration.

\pagebreak
%%%%%%%%%%%%%%%%%
\bibliography{bibi} %%% ref.bib file
%%%%%%%%%%%%%%%%%

\end{document}